\documentclass[twocolumn, tighten, twocolappendix]{aastex63}

\usepackage{times}
\usepackage{amsmath}
\usepackage{bm}
\usepackage{graphicx}
\DeclareGraphicsExtensions{.pdf,.eps,.png,.jpg}
\usepackage{acronym}
\usepackage{color}
\usepackage{subfigure}
\usepackage{tabularx}
\newcolumntype{Y}{>{\centering\arraybackslash}X}
\usepackage{longtable}
\usepackage{multirow}
\usepackage{dcolumn}
\usepackage{soul}
\usepackage{float}
\usepackage[figuresleft]{rotating}

\allowdisplaybreaks

\newcommand{\msun}{\mbox{$M_\odot$}}

\shortauthors{Kim et al.}

\begin{document}

\title{Identification of Lensed Gravitational Waves with Deep Learning}

\author[0000-0003-1653-3795]{Kyungmin~\surname{Kim}}
\email{kkim@kasi.re.kr}
\affiliation{Korea Astronomy and Space Science Institute, 776 Daedeokdae-ro, Yuseong-gu, Daejeon 34055, South Korea}
\affiliation{Department of Physics, Ewha Womans University, 52 Ewhayeodae-gil, Seodaemun-gu, Seoul 03760, South Korea}

\author[0000-0003-2510-4132]{Joongoo~\surname{Lee}}
\affiliation{Korea Astronomy and Space Science Institute, 776 Daedeokdae-ro, Yuseong-gu, Daejeon 34055, South Korea}
\affiliation{Department of Physics and Astronomy, Seoul National University, Seoul 08826, South Korea}

\author{Robin~S.~H.~\surname{Yuen}}
\affiliation{Department of Physics, The Chinese University of Hong Kong, Shatin, New Territories, Hong Kong}

\author[0000-0002-3887-7137]{Otto~A.~\surname{Hannuksela}}
\affiliation{Nikhef - National Institute for Subatomic Physics, Science Park, 1098 XG Amsterdam, The Netherlands}
\affiliation{Department of Physics, Utrecht University, Princetonplein 1, 3584 CC Utrecht, The Netherlands}

\author[0000-0003-4297-7365]{Tjonnie~G.~F.~\surname{Li}}
\affiliation{Department of Physics, The Chinese University of Hong Kong, Shatin, New Territories, Hong Kong}

\begin{abstract}
Similar to light, gravitational waves (GWs) can be lensed. Such lensing phenomena can magnify the waves, create multiple images observable as repeated events, and superpose several waveforms together, inducing potentially discernible patterns on the waves. In particular, when the lens is small, $\lesssim 10^5 \msun$, it can produce lensed images with time delays shorter than the typical gravitational-wave signal length that conspire together to form ``beating patterns''. We present a proof-of-principle study utilizing deep learning for identification of such a lensing signature. We bring the excellence of state-of-the-art deep learning models at recognizing foreground objects from background noises to identifying lensed GWs from noise present spectrograms.
We assume the lens mass is around $10^3 \msun$ -- $10^5 \msun$, which can produce the order of millisecond time delays between two images of lensed GWs. We discuss the feasibility of distinguishing lensed GWs from unlensed ones and estimating physical and lensing parameters. Suggested method may be of interest to the study of more complicated lensing configurations for which we do not have accurate waveform templates.
\end{abstract}

\keywords{Gravitational waves (678); Weak gravitational lensing (1797); Astronomy data analysis (1858); Convolutional neural networks (1938)}


\section{Introduction}

When gravitational waves (GWs) propagate near massive astrophysical objects such as black holes, substructures, galaxies, or galaxy clusters, they can be gravitationally lensed~\citep{Ohanian:1974, Bliokh:1975apss, Bontz:1981apss, Thorne:1983, Deguchi:1986apj, Schneider:1992, Nakamura:1999ptps, takahashi:2003apj}. 
Gravitational lensing is verified by electromagnetic (EM) observations for decades and has led to several groundbreaking findings in astrophysics. 
For example, it has resulted in the detection of exoplanets~\citep{cassan2012one} and has placed highly credible evidence for dark matter~\citep{clowe2004weak,markevitch2004direct}. 
Similarly, lensing of GW is also expected to present interesting applications in fundamental physics, astrophysics and cosmology as studied in the literature~\citep{2011MNRAS.415.2773S, Collett:2016dey, Baker:2016reh, Fan:2016swi, Liao:2017ioi, Lai:2018prd, dai2018detecting, Jung:2017flg, Li:2019rns, Cao:2019kgn, Hannuksela:2020xor}.

The methods to detect lensed GWs have been developed in recent years, and the first searches for GW lensing signatures in the LIGO and Virgo data were carried out recently~\citep{Hannuksela:2019kle,Li:2019osa,McIsaac:2019use,Pang:2020qow,Dai:2020tpj}. 
Based on the predictions on the number of expected GW sources and the distribution of lenses in the Universe, the current forecasts on the lensing rate suggest that the Advanced LIGO~\citep{aLIGO} and Virgo~\citep{AdV} detectors detect strongly lensed GWs upon reaching design sensitivity~\citep{KenKaze:2018prd, li2018gravitational,Oguri:2018muv}.\footnote{In particular, \citet{KenKaze:2018prd,li2018gravitational, Oguri:2018muv} estimated around $\mathcal{O}(1)$ lensed event per year at the design sensitivity.}
The rate of lensing will significantly increase further when the Einstein Telescope~\citep{Punturo_2010} or Cosmic Explorer~\citep{Reitze:2019iox} come online; under the design sensitivity of these detectors, microlensing might also become more prominent~\citep{Diego:2019lcd}. Therefore, it is timely to explore all promising candidate methods which will be a standard for searching lensed GW signals.

In this work, our interest is focused on GWs from compact binary mergers because all observed GWs reported in GWTC-2~\citep{Abbott:2020niy} originated from the mergers. To date, such GWs have been targeted by the template-based searches such as \texttt{PyCBC}~\citep{alex_nitz_2020_4075326, Usman:2015kfa} and \texttt{gstlal}~\citep{Sachdev:2019vvd, Messick:2016aqy}. For searching lensed GWs, only limited template-based methods have been suggested: \citet{Lai:2018prd, Hannuksela:2019kle}, for example, discussed the template-based search method under assuming isolated point mass lenses. Nonetheless, there are no complete waveform templates for the lensed GWs yet, including the microlensed GWs, that are currently applicable to LIGO/Virgo searches for the identification of lensing-induced beating patterns. Moreover, it is conceivable that more complicated lens setups such as a field of microlenses within an external potential also cause the beating patterns~\citep{Diego:2019lcd, Pagano:2020rwj, Diego:2019rzc, Cheung:2020okf} and building a generalized complete template model for the setup will be much harder. Therefore, it is important to explore alternatives to the traditional templated searches.

On the other hand, for the recent decade, applying machine learning (ML) techniques to various disciplines of GW science has been intensively studied~\citep[for a comprehensive review, see][and references therein]{Cuoco_2020}. From those studies, it turns out that ML is promising in improving data quality, enhancing search performance, and accelerating analysis speed. With considering such advantages, \citet{Singh:2018} discussed an alternative method utilizing ML for the discrimination of lensed GWs. This study showed that the lensing-induced superposition of GW signals could be distinguished from unlensed GWs via supervised ML classifiers with $>90\%$ accuracy.

In~\citet{Singh:2018}, we configured the input data with the spectrogram of simulated signals of lensed GWs to facilitate the ML-based classification. The beating pattern caused by the lensing effect can be seen in either time-domain or frequency-domain GW data. However, from the spectrogram, we can read out not only the beating pattern in time-domain but also the frequency-dependent amplification simultaneously. It is also known that spectrogram is advantageous in using convolutional neural networks (CNN) for capturing glitches --- non-stationary and non-Gaussian transient noises presented with complex morphologies --- in GW data \citep{George:2017fbn, George:2017pmj}. We draw analogies between the identifications of glitch and lensed GW because both can be considered as unmodeled analysis in lieu of accurate templates. Moreover, different lensing configurations vary the lensing signature even for a single GW signal from a compact binary merger. This makes the identification of lensing in GWs difficult in the absence of a complete lensed waveform model. Therefore, the use of spectrogram is beneficial in the ML-based identification of the lensing signature in consideration than the use of either time-domain or frequency-domain data only.

In this paper, we present a proof-of-principle study extending the work of \citet{Singh:2018}: we elaborate the previous approach by taking into account more realistic considerations such as (i) using a realistic GW waveform generation method, (ii) adopting a power spectral density (PSD) model of the Advanced LIGO, (iii) constraining the optimal signal-to-noise ratio (SNR) to cover the SNRs of the GW signals from binary black hole mergers (BBH) which are reported in the GWTC-2, and (iv) using a conventional constant-Q transform method~\citep{Chatterji:2004qg} in the generation of spectrogram samples. For the application of machine learning, we use a state-of-the-art deep learning model, the Visual Geometry Group network (VGG)~\citep{Simonyan14c} --- a variant of typical CNN. The implementation of the VGG is conducted not only for the classification but also for the regression on the characteristic parameters of the lensed GWs. From the performance test of regression, we find that the matches between the distributions of the true and predicted parameters are $\gtrsim 0.82$ and the residual between the true and predicted values of densely sampled parameters are mostly placed within the $1\sigma$ confidence interval of the distribution of residuals. We also find that $\gtrsim 97\%$ of lensed GW samples are correctly classified from the performance test of classification.

This paper is organized as follows: we introduce a brief overview of considered lensing models in Sec.~\ref{sec:lens}. We summarize the procedure from the preparation of input data to the application of VGG in Sec.~\ref{sec:method}. 
We present the result of performance tests of regression and classification in Sec.~\ref{sec:results}. We also provide the result of a simple validation test on another set of spectrograms prepared with different parameter population models from~\cite{Abbott:2020gyp} in the same section. Finally, we discuss this work and future direction in Sec.~\ref{sec:discussion}.


\section{Lensed Gravitational Waves}
\label{sec:lens}

Here we briefly recap the two standard analytical gravitational lens models in the context of gravitational-wave lensing -- the point mass lens (PM) and the singular isothermal sphere model (SIS). For a more thorough review, we refer the reader to, e.g.,~\citet{takahashi:2003apj}.

Depending on the wavelength, $\lambda$, of the GW, the optics describing lensing effect can be categorized in two regimes: the geometrical optics limit is valid when $\lambda$ is much shorter than the Schwarzschild radius of a lens. On the other hand, when $\lambda$ is larger than the Schwarzschild radius of a lens, the diffraction effect becomes important, and the wave optics limit holds for the regime \citep{takahashi:2003apj}. The regime where each optics limit is valid also can be determined with respect to $M_L$. For example, it is discussed in~\citet{takahashi:2003apj} that the diffraction effect becomes important when $M_L$ satisfies 
\begin{equation}
M_{L} \lesssim 10^8 \msun \left( f / \mathrm{mHz} \right)^{-1}. \label{eqn:diffraction_ML}
\end{equation}
If we regard the sensitive frequency band, $10^2 \mathrm{Hz}$ -- $10^3 \mathrm{Hz}$, of the Advanced LIGO and Virgo detectors, we can easily compute from Equation~\eqref{eqn:diffraction_ML} that $M_L$ in the range of $10^2\msun$ -- $10^3\msun$ corresponds to the regime where the wave optics limit is valid. However, for the 10 GWs from the BBH in GWTC-1, no evidence of micro/millilensing by isolated point mass lenses for $M_L \lesssim 10^5 \msun$~\citep{Hannuksela:2019kle} was found. Another study~\citep{Lai:2018prd} showed that the wave optics effect is typically negligible when $M_L$ is $\sim 10^3\msun - 10^4 \msun$. Therefore, we focus on the geometrical optics limit only in this work for simplicity.

\begin{figure}[t!]
\centering
\includegraphics[width=.95\linewidth]{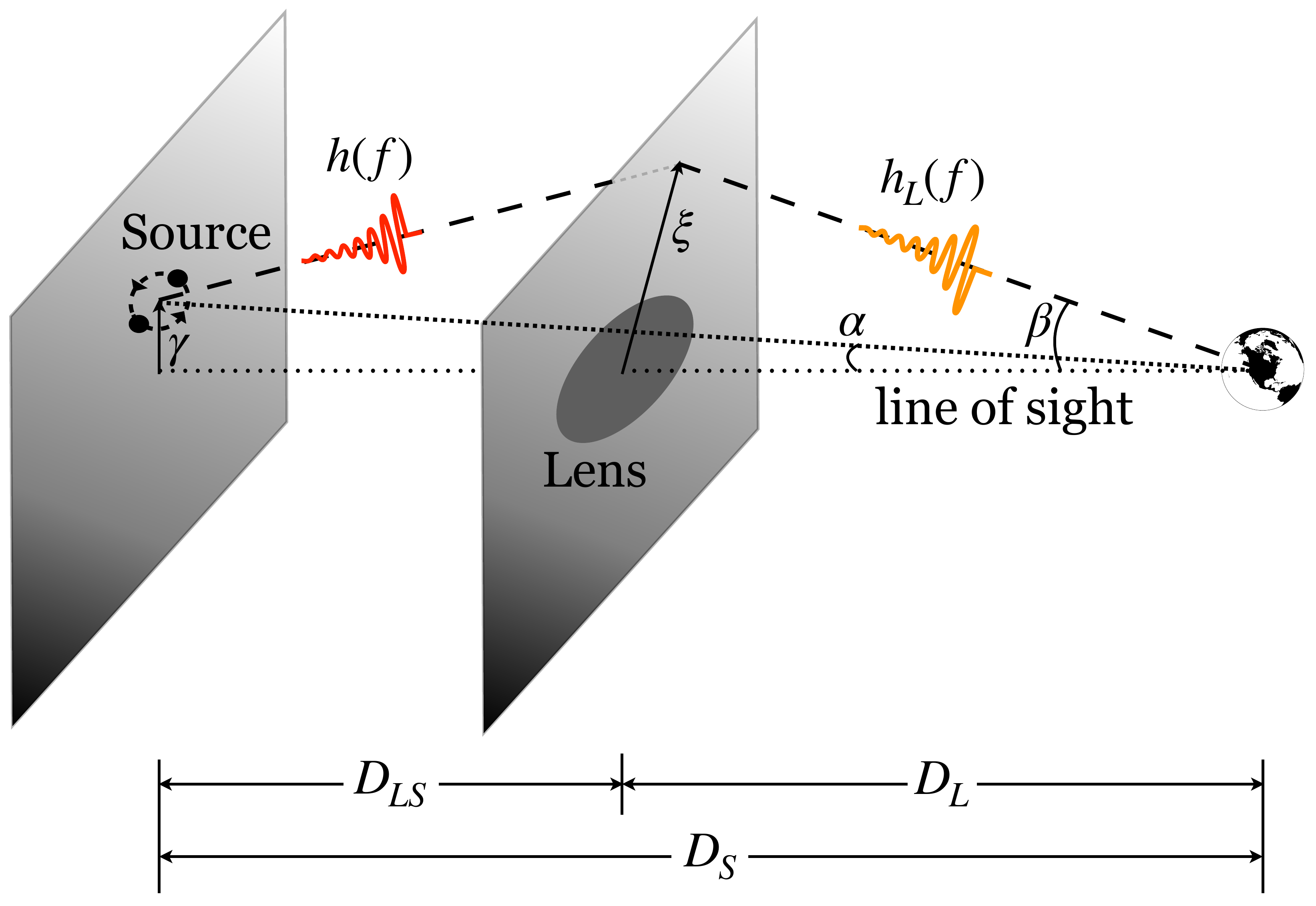}
\caption{Schematic configuration of GW lensing in thin lens approximation where the angle, $\alpha$ and $\beta$ are negligible. $\bm{\gamma}$ is the position of the source in the source plane which measures the displacement of the position of the source from the line of sight, $\xi$ is the impact factor which can be normalized by the Einstein radius, $\xi_0$. $D_{L}$, $D_{S}$, and $D_{LS}$ indicate the distances from observer to the lens, to the source, and from the lens to the source, respectively. The differences in colors on $h(f)$ and $h_{L}(f)$ represent the frequency-dependent amplification effect in GW described in Equation~(\ref{eq:amplification}).}
\label{fig:lensmodel}
\end{figure}

We assume the thin lens approximation where $M_L$ is distributed over a two-dimensional plane perpendicular to the observer's line of sight to the lens; with the considered lens models, PM and SIS, we can specify the surface mass density of a lens. In Figure~\ref{fig:lensmodel}, we illustrate a schematic configuration of the lensing of GW in the thin lens approximation. By the approximation, we can parameterize the displacement of source, $\gamma$, and the impact factor, $\xi$, with a single value, $y$, such as
\begin{equation}
y = \frac{\gamma D_{L}} {\xi_0 D_{S} }, \label{eq:y}
\end{equation} 
where $\xi_0 = \sqrt{(4 G M_L / c^2) D_{LS} D_L / D_S}$ is the Einstein radius of a lens; $D_L$, $D_S$, and $D_{LS}$ are the distances to the lens, to the source, and from the lens to the source, respectively. With the above setup, the relation between the lensed GW, ${h_{L}}(f)$, and the unlensed GW, $h(f)$ in frequency domain can be written as~
\begin{equation}
h_{L}(f) = F(f) h(f), \label{eq:amplification}
\end{equation}
where $F(f)$ is the amplification factor which can be determined by $y$ and the surface mass density of lens. 

In the geometrical optics limit, $F(f)$ of each model is written as
\begin{eqnarray}
\textrm{PM}: & F(f) &= \sqrt{|\mu_{+}|} - i \sqrt{|\mu_{-}|} e^{2\pi i f \Delta t_d} \label{eq:F(f)_pm}\\
\textrm{SIS}: & F(f) &= \begin{cases}
    \sqrt{|\mu_{+}|} - i \sqrt{|\mu_{-}|} e^{2\pi i f \Delta t_d}, & y \leq 1,\\
    \sqrt{|\mu_{+}|},  & y \geq 1, \label{eq:F(f)_sis}
\end{cases}
\label{eq:F(f)}
\end{eqnarray}
where $\mu_\pm$ and $\Delta t_d$ are the magnification factors of two images and the time delay between them, respectively. 
$\mu_\pm$ and $\Delta t_d$ are respectively given by 
\begin{eqnarray}
\mu_\pm &=& \frac{1}{2} \pm \frac{y^2 + 2}{2y\sqrt{y^2 + 4}}, \label{eq:mu_pm}\\
\Delta t_d &=& \frac{4 G M_{Lz}}{c^3} \left[ \frac{y \sqrt{y^2 + 4}}{2} + \ln \left\{ \frac{\sqrt{y^2+4}+y}{\sqrt{y^2+4}-y} \right\} \right], \label{eq:dt_pm}
\end{eqnarray}
for the PM and
\begin{eqnarray} 
\mu_\pm &=& \pm1 + \frac{1}{y}, \label{eq:mu_sis}\\
\Delta t_d &=& \frac{8 G M_{Lz} y}{c^3}, \label{eq:dt_sis}
\end{eqnarray}
for the SIS. In Equations~\eqref{eq:dt_pm} and \eqref{eq:dt_sis}, $M_{Lz} = M_L (1+z_L)$ is the redshifted mass of a lens. From Equations~\eqref{eq:mu_pm} -- \eqref{eq:dt_sis}, one can see that $\mu_\pm$ and $\Delta t_d$ are given as the function of $y$ and $M_{Lz}$. In this work, we regard $y$ as free parameter and will discuss the constraint on the range of its value in the following section. 

As noted earlier, it is possible to consider more complex lensing configurations such as microlenses embedded in an external potential. The lensing models we have chosen are simple (analytical solutions are accessible) but sufficient to produce qualitative lensing features, such as beating patterns, in GW waveforms inherent to more complex configurations (see~\citet{Pagano:2020rwj}).\footnote{Another mature method enabling more complicated lensing configurations, including wave optics effects, is under construction~\citep{Cheung:2020okf}.}


\section{Method}
\label{sec:method}

\subsection{Data Preparation}
\label{sec:data_prep}

This work's main goal is to identify lensed GWs with deep learning from the confirmed detections.\footnote{See Figure~10 in~\citet{LIGOScientific:2018mvr} for the illustration of a similar procedure using the standard nested sampling analysis} To this end, we describe the details of data preparation from the generation of simulated GWs to the preparation of practical spectrogram samples for the purpose.

\subsubsection{Gravitational-Wave Model}
\label{sec:GWmodel}

We firstly consider a waveform model for GWs from the BBH. Specifically, we mainly focus on the inspiral phase instead of the merger and ringdown phases. The portion of the latter two phases is much shorter than the inspiral phase to see the superposed waveform due to the two images of a lensed GW arrived at different times with $\Delta t_d$. In~\citet{Singh:2018}, we have indeed shown that considering only the inspiral phase without the precision effect of the BBH system is sufficient to study the feasibility of the classification of lensed GWs from unlensed ones even in the pedagogical approach. 

On the other hand, a precessing binary system originating from the non-zero and misaligned component spins can introduce modulation in the amplitude and phase of GWs~\citep{Hannam:2013pra}. It may appear similar to the beating pattern due to the superposition of two non-precessing GWs. Therefore, we also need to examine our ability to discriminate lensed GWs from the precessing ones. However, in this work, we discard the lensing of precessing GWs for simplicity.

We use the IMRPhenomPv2 model~\citep{Schmidt:2014iyl,Hannam:2013oca} for the generation of simulated waveforms of both non-processing and precessing GWs in a consistent manner. In the use of the waveform model, we set the range of the mass of individual component objects as $4 \msun$ -- $35 \msun$ which is commonly used for searching GWs from the BBH merger in the conventional template-based GW data analysis methods such as \texttt{PyCBC}~\citep{alex_nitz_2020_4075326, Usman:2015kfa} and \texttt{gstlal}~\citep{Sachdev:2019vvd, Messick:2016aqy}. For the processing GWs, we set the spin of the component objects as $s_1, s_2 \in [-1,1]$.

\begin{figure}[t]
\centering
\includegraphics[width=1.\linewidth]{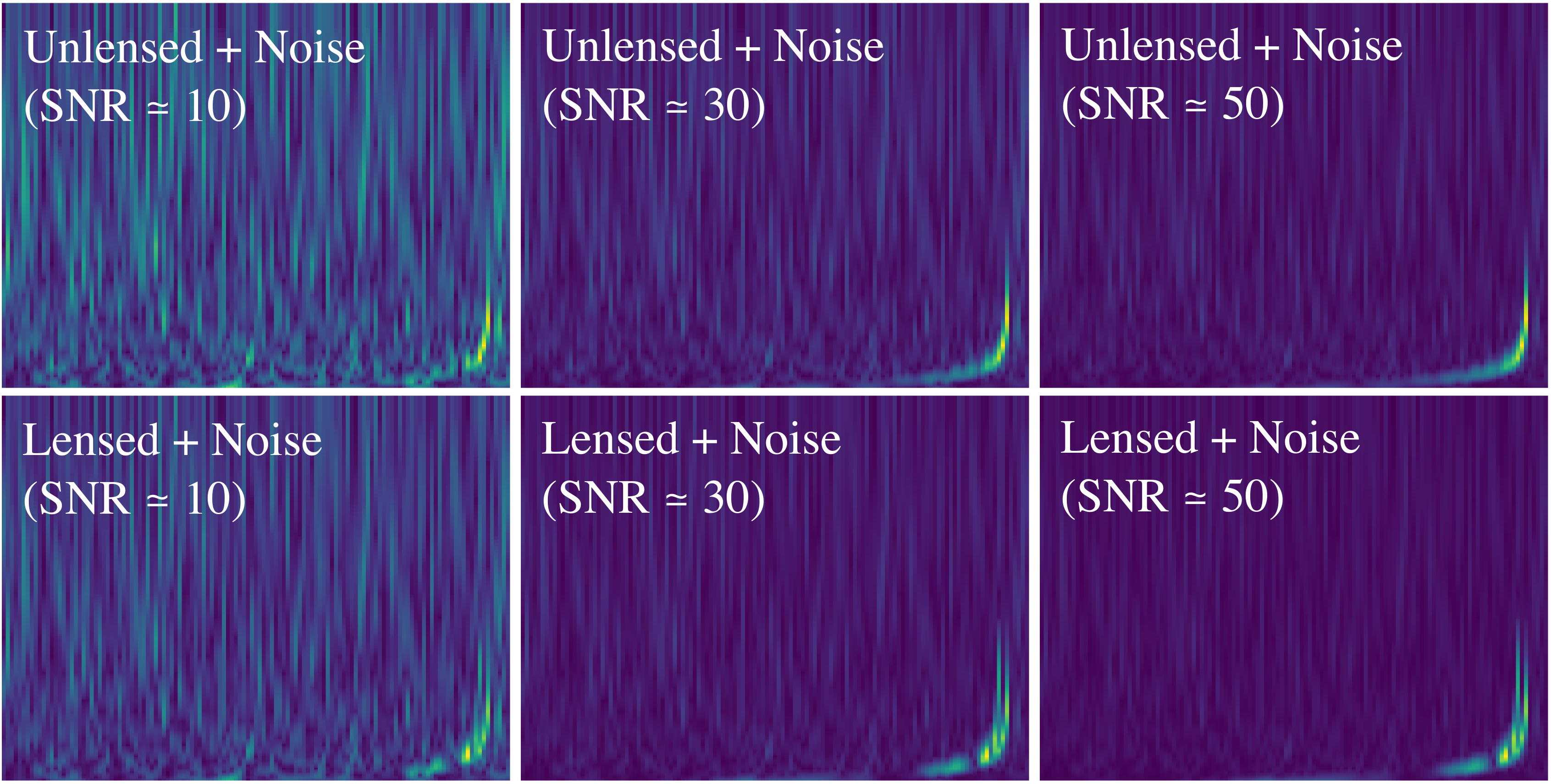}
\caption{Illustrative examples of the spectrogram of an unlensed GW signal (top) and its lensed counterpart (bottom) in the presence of different levels of noise, 10, 30, and 50 (from left to right). The chirp signals are generated by using the same parameters, $m_1 = m_2 = 20 \msun$ and $D_S = 1$ Gpc, for the source system. For the lensed ones, we use PM model with $D_L = 800$~Mpc and $M_L = 10^4 \msun$. Here, we artificially rescale the power spectral density of considered noise model to produce the annotated values of SNR for a heuristic purpose. One can see that, as SNR increases from 10 to 50, the level of discrepancy between lensed and unlensed signals that appeared in the bottom-right corner of the spectrogram increases too.
\label{fig:spectrogram_example}}
\end{figure}

\subsubsection{Constraint on Position Parameter}

Since we are interested in the lensing effect imprinted in GW, especially the beating pattern in the spectrogram, the order of time delay, $\Delta t_d$, between two images of a lensed GW is expected to be the order of millisecond. From Equations~\eqref{eq:dt_pm} and \eqref{eq:dt_sis}, we could see that $\Delta t_d$ depends not only on $M_L$ but also on $y$. Hence, we need to consider an appropriate range for $y$ that can induce such a time delay. We can understand from Equation~\eqref{eq:F(f)_sis} that, when $y \geq 1$, $F(f)$ is determined by a single magnification factor, that is, only by $\mu_{+}$ without $\mu_{-}$. It means that only a single amplified image is formed around the lens in this case. Thus, to produce the observable beating pattern caused by two superposed images of a lensed GW, we limit the range of $y$ to satisfy $0.05 < y < 1$ for both PM and SIS models.

\subsubsection{Preparation of Realistic Spectrogram Samples}
\label{sec:spectrogram_prep}

In general, real GW data of LIGO/Virgo detectors contain not only GW signals but nonnegligible noises originating from various stationary and non-stationary noise sources which may hinder the identification. We suppose that noise level represented in the spectrogram is critical in the ML-based identification of lensed GWs in the situation. For a simple examination of the supposition's validity, we present illustrative example spectrograms in Figure~\ref{fig:spectrogram_example}: When SNR is $\sim10$ (left panels), the relatively higher noise level makes indistinguishable spectrograms. Meanwhile, we can quickly distinguish the lensed and unlensed spectrograms as SNR increases to $\sim30$ and $\sim50$ (middle and right panels, respectively) for the same chirp signal. Therefore, to study the capability of identifying lensed GW signals from noisy data, we adopt a theoretical PSD model, the Detuned High Power model of the Advanced LIGO~\citep{ZeroDetHighP}, and prepare spectrograms imitating the real scenario. We add the noise data to the simulated signals described in the previous section.

We constraint the SNRs of spectrogram samples to satisfy $10 \leq$ SNR\footnote{The SNR computed in this work is the optimal SNR because the noise and the exact form of GW signal are known~\citep{Jolien_book}.} $\leq 50$, covering the SNRs of the BBH-GW signals reported in the GWTC-2. Note that actual SNRs reported in GWTC-2 are in the range of $10 \leq \mathrm{SNR} \leq 24.4$. However, it is estimated that we may obtain SNR $\sim$ 50 for a GW signal produced from a BBH merger of $m_1 = m_2 = 30 \msun$ at $z=10$ even under a pessimistic assumption on the design sensitivity of the third generation GW detectors~\citep{Evans:2016mbw}. Therefore, regarding the design sensitivity of the future detectors together with the SNRs in GWTC-2, we set the upper bound of SNR about two times higher than the highest SNR in GWTC-2.
Further, we apply a conventional time-frequency representation called the constant-Q transform~\citep{Chatterji:2004qg} to generate our spectrogram samples. 

In Table~\ref{tab:GWparams}, we summarize the parameters used in the preparation of the final form of spectrogram samples reflecting the considerations and constraints hitherto discussed. The source and lens masses are sampled from a logarithmic distribution to reduce the bias towards heavier masses. Other parameters are sampled from uniform distribution within the given range for simplicity.\footnote{Note that considering adequate distributions on the parameters is hard at the moment since we do not have any complete model for the astrophysical objects in consideration yet. Although the GWTC-2 provides more knowledge on the GWs’ progenitors, e.g.,~\cite{Abbott:2020gyp}, the information might be insufficient to build complete models because it has come from the detection of only $\sim$50 events. While investigating more appropriate black hole and gravitational lens populations can be a worthwhile long-term endeavor, we would need much more complex lensed waveforms to perform rigorous population studies, including wave optics effects, macro models, and micro/millilens fields (for a recent attempt, one can refer to~\citet{Cheung:2020okf}.). Therefore, assuming the arbitrary and simpler distribution is an inevitable choice for this proof-of-principle study in the absence of such complete models.}

\begin{table}[t]
\caption{Our prior choices for the lensed GW parameters. We apply the same parameter ranges to the generation of both nonprecessing and precessing waveforms. For the precessing waveforms, the range of component spins is applied for x-, y-, and z-directions individually. }
\begin{center}
\begin{tabularx}{1\linewidth}{X c}
\toprule
Parameter & Range \\
\hline
Component masses of source, $m_1$ \& $m_2$ & $4$ -- $35\msun$ \\
Lens mass, $M_L$ & $10^3$ -- $10^5\msun$ \\
Distance to lens, $D_L$ & $10$ -- $10^3$ Mpc \\ 
Distance from lens to source, $D_{LS}$ & $10$ -- $10^3$ Mpc \\
Displacement of source from the line of sight, $\gamma$ & $0$ -- $0.5$ pc \\
Redshift, $z$ & $0$ -- $2$ \\
Signal-to-noise ratio, SNR & $10$ -- $50$ \\
\hline
Component spins, $s_1$ \& $s_2$ & [-1,~1]\\
\hline
\hline
\end{tabularx}
\end{center}
\label{tab:GWparams}
\end{table}

The number of resulting spectrogram samples is 45,000 for each of (i) lensed ($L$), (ii) unlensed and non-processing ($U_N$), and (iii) unlensed and precessing ($U_P$) GWs.

\subsection{Application of Deep Learning}
\label{sec:deeplearning}

This section presents a brief introduction of the deep learning model adopted in this work and the detailed procedure of training the deep learning model. 

\subsubsection{Visual Geometry Group Network}

The VGG network structure has been designed as a variant of the classical type of convolutional neural networks, which is inspired by the AlexNet~\citep{NIPS2012_4824}. The central insight of the VGG is using only 3x3 convolutional filters in the construction of deeper networks.\footnote{There is a competitive model called Inception~\citep{Inception_v1,Inception_v2v3,Inception_v4} also known as GoogLeNet. It is shown that the VGG is more robust for face (foreground) recognition in the presence of noise (background) than Inception~\citep{Grm_2018}. Thus, we decide to use VGG since we are also interested in recognizing lensed GW signals (foreground) from the presence of noise (background).} 

For the classification and regression of our interest, we use the latest version of the VGG, VGG-19, and implement it with \texttt{PyTorch}~\citep{NEURIPS2019_9015}. We summarize the structural details used for the implementation of VGG-19 in Appendix~\ref{apx:vgg}. Hereafter we refer to the VGG-19 as VGG.

\subsubsection{Input Data}
\label{sec:input_data}

For the application of the VGG to the spectrogram samples prepared in Section~\ref{sec:spectrogram_prep}, we split the samples into three subset data such as training, development, and evaluation data. We randomly choose 80\% of total samples for the training data and 10\% each for the development and evaluation data.

We then perform a preprocessing, the min-max normalization, on the samples in all subsets by rescaling pixels' values in a spectrogram sample to be normalized between $-1$ and 1. It is known that this preprocessing helps to mitigate the vanishing gradient problem~\citep{Kolen:01book}, which could be occurred in the training. Another benefit of this preprocessing is that we can keep the information of SNR of an original sample even with the normalization. We use each of the preprocessed subset data as the input data fed into each step of training, development, and evaluation, respectively.

Meanwhile, particularly for the regression, we prepare additional data called target data with the parameters to be predicted by the VGG on top of the spectrogram samples. For the parameters, we consider $M_{S}^{c}$, $M_{L}$, $z_{S}$, and $z_{L}$. Each of the parameters is standardized with the min-max normalization since (i) the ranges of parameters to be predicted by the VGG are quite different as shown in Table~\ref{tab:GWparams} and (ii) diversely ranged target data may cause biased results. Consequently, we can restrict the contribution of each target value into a comparable range and may avoid the diversely ranged target parameters contributing differently to the training. Finally, we conduct a postprocessing on the predicted outputs for the final estimation of the parameters by reversely performing the min-max normalization to recover their original value.

\subsubsection{Training VGG}
\label{sec:training}

We set a batch size to be 128 for both classification and regression. The maximum training epoch is set as 100 while iterating whole samples in the training data. To boost the training and enhance the accuracy of the VGG, we use the Adam optimization algorithm~\citep{2014ADAM}, one of the extensions of the stochastic gradient descent method. For the implementation of the learning rate decay, we reduce the learning rate by a factor of $2$ from the initial learning rate when the training stops updating the VGG since it is known that most deep learning models often benefit from the reduction in the learning rate. During the training, we keep tracking the validity of a temporal model at each epoch and save a checkpoint for the temporal model to check improvement compared to the former best performing model on the same validation data. The training is conducted on NVIDIA Tesla P40 GPU with 24GB memory. We summarize the above setups in Table~\ref{tab:TrainDetail}.

\begin{table}[b]
\caption{The training setup of the VGG model. We employ the configurations for both classification and regression.}
\begin{center}
\begin{tabularx}{1\linewidth}{X c}
\toprule
Parameter & Value \\
\hline
Batch size & 128 \\
Maximum training epoch & 100 \\
Optimization algorithm & Adam \\
Learning rate decay strategy & On plateau \\
Initial learning rate & 8e-5 \\
Minimum learning rate & 1e-5 \\
Decay factor & 2 \\
Computing processor & NVIDIA Tesla P40 (24GB) \\
\hline
\hline
\end{tabularx}
\end{center}
\label{tab:TrainDetail}
\end{table}

We identically apply the training setup for both classification and regression. However, we alter the form of output for each task such that let the classification return the probability to each sample type either $U_N$, $U_P$, or $L$ via the one-hot encoding and let the regression return the normalized value of parameters. The loss functions adopted for the error measurement of the training of each task are summarized in Appendix~\ref{apx:loss_func}.

\begin{figure*}[t]
    \centering
    \subfigure[$L_\mathrm{PM}$ - $M_S^{c}$]
    {
        \includegraphics[width=.4\linewidth]{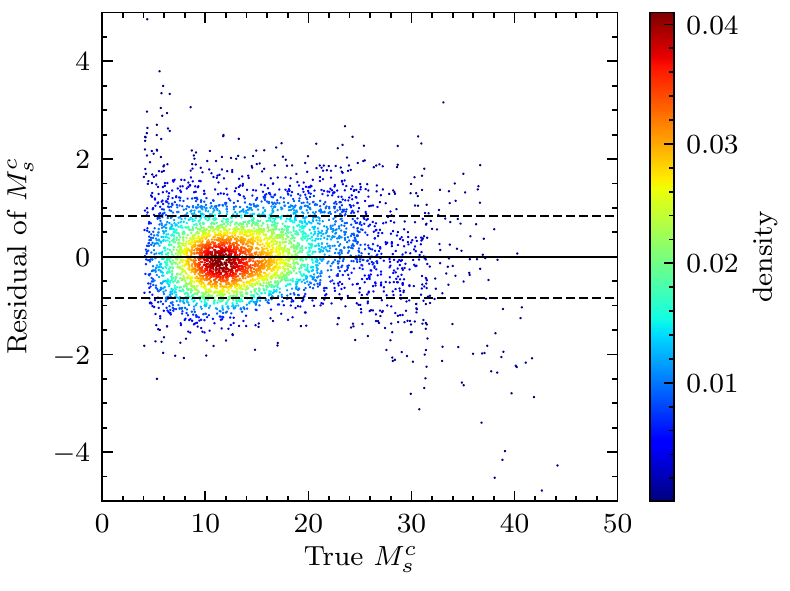}
    }
    \subfigure[$L_\mathrm{PM}$ - $M_L$]
    {
        \includegraphics[width=.4\linewidth]{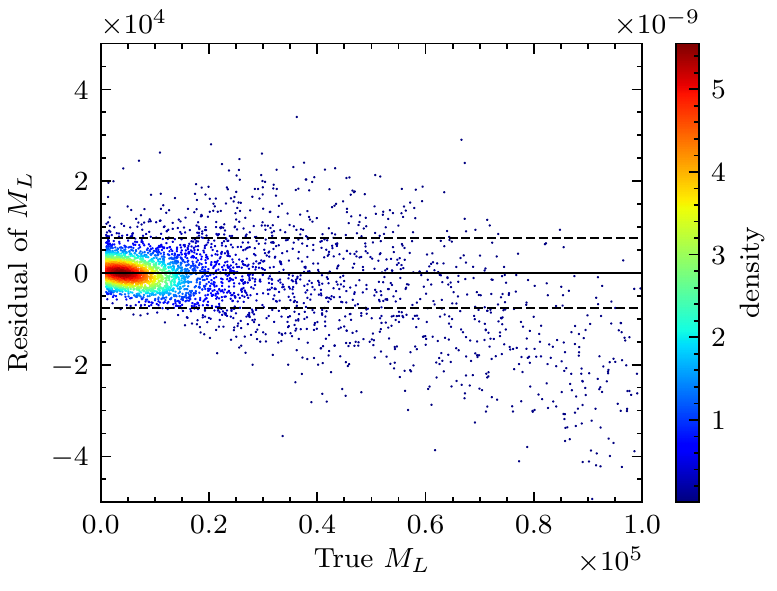}
    }\\
    \subfigure[$L_\mathrm{PM}$ - $z_S$]
    {
        \includegraphics[width=.4\linewidth]{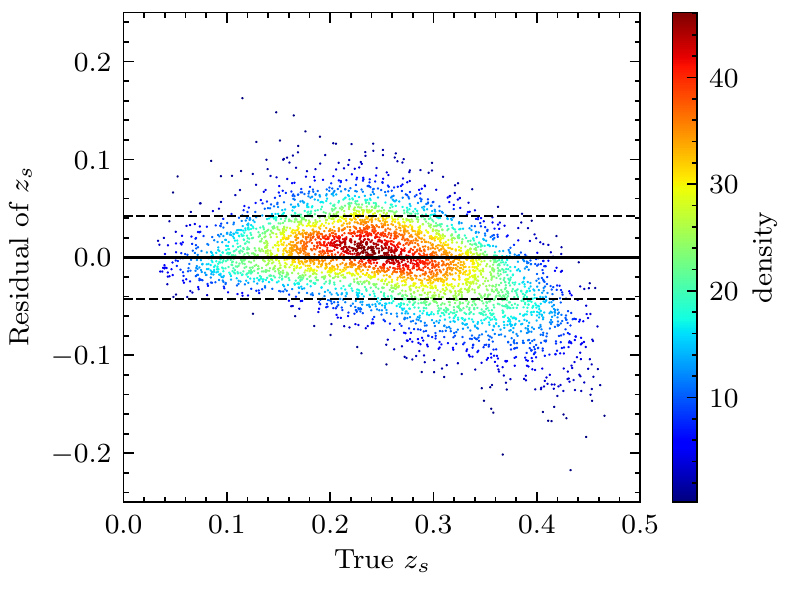}
    }
    \subfigure[$L_\mathrm{PM}$ - $z_L$]
    {
        \includegraphics[width=.4\linewidth]{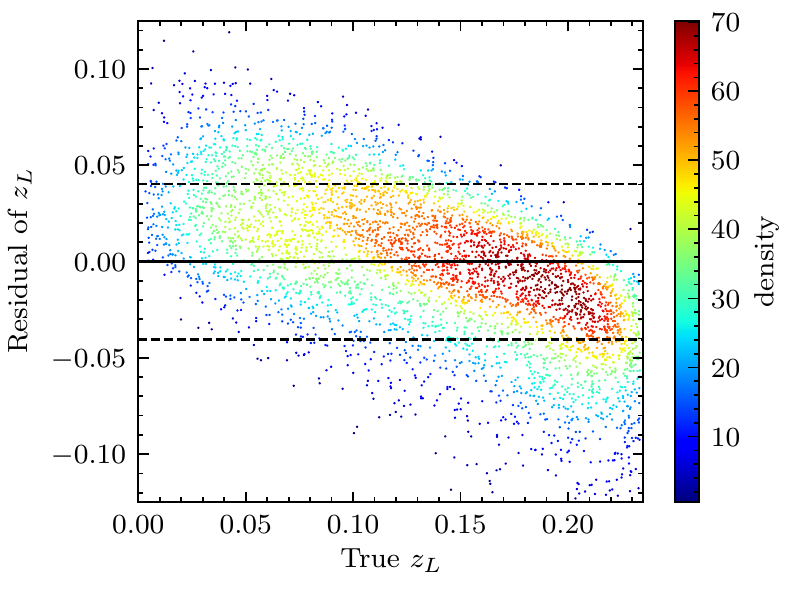}
    }
    \subfigure[$L_\mathrm{PM}$ - $y$]
    {
        \includegraphics[width=.4\linewidth]{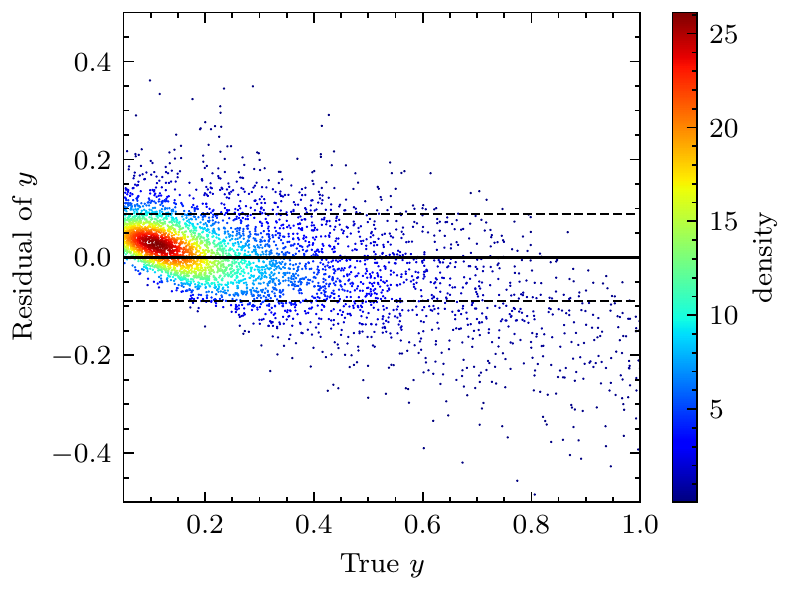}
    }
    \caption{Residual plots of $L_\mathrm{PM}$. The x-axes indicate the true value of parameters of $L_\mathrm{PM}$; the y-axes indicate residual of the predicted values against the true values of the parameters. The horizontal dashed lines represent the $1\sigma$ confidence interval of each of the residuals. The units of $M_S^c$ and $M_L$ are in $M_\odot$; others are dimensionless. The residuals to a densely populated range of true parameter mostly distribute within the 1$\sigma$ confidence interval. Meanwhile, the residuals to a sparsely populated range of true parameters tends to either overestimated prediction (points over the upper bound of 1$\sigma$) or underestimated prediction (data points under the lower bound of 1$\sigma$).
    \label{fig:reg_residual_pm}}
\end{figure*}

\begin{figure*}[t!]
    \centering
    \subfigure[$L_\mathrm{PM}$ - $M_S^{c}$]
    {
        \includegraphics[width=.4\linewidth]{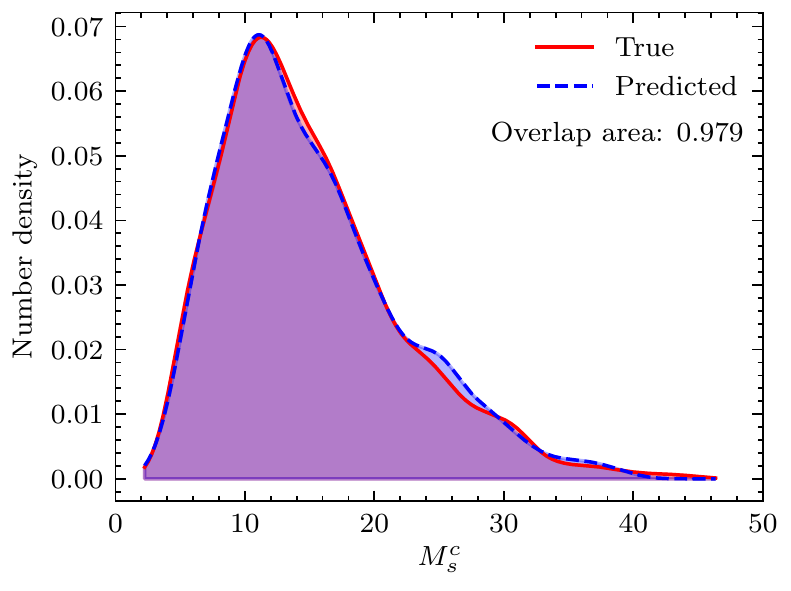}
    }
    \subfigure[$L_\mathrm{PM}$ - $M_L$]
    {
        \includegraphics[width=.4\linewidth]{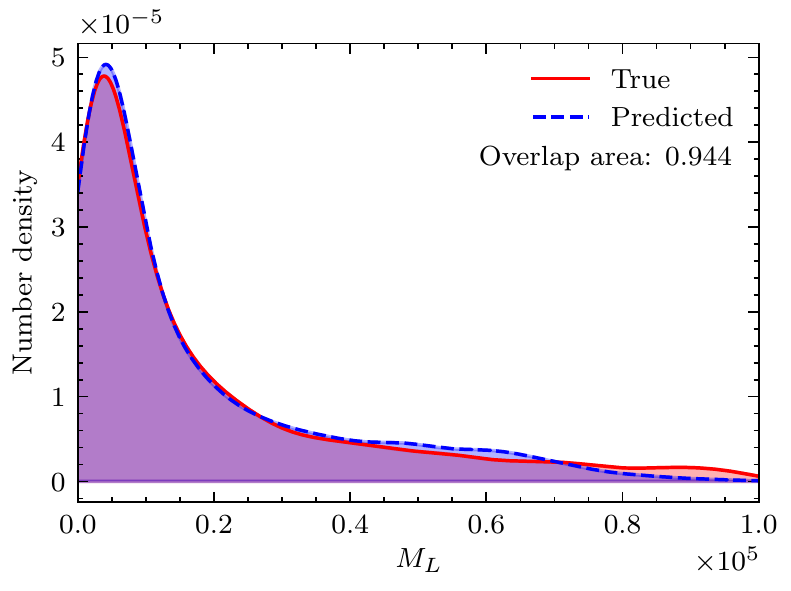}
    }\\
    \subfigure[$L_\mathrm{PM}$ - $z_S$]
    {
        \includegraphics[width=.4\linewidth]{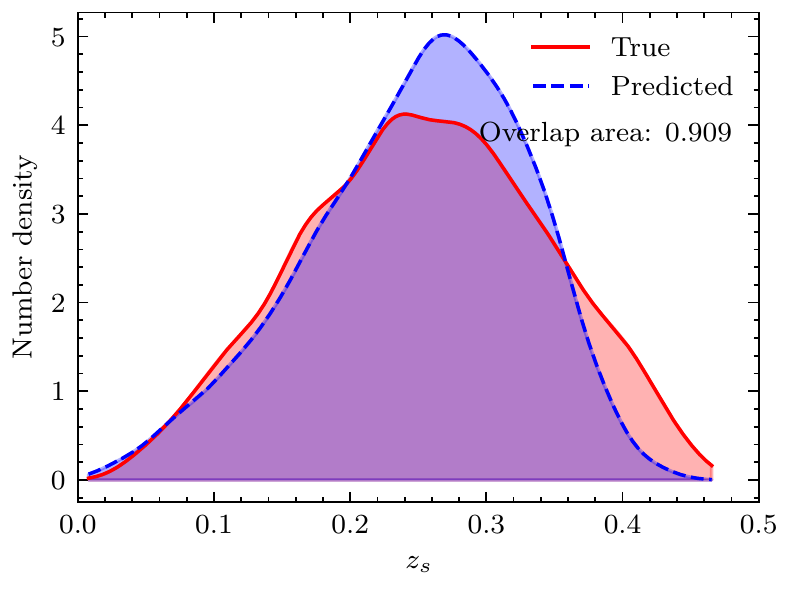}
    }
    \subfigure[$L_\mathrm{PM}$ - $z_L$]
    {
        \includegraphics[width=.4\linewidth]{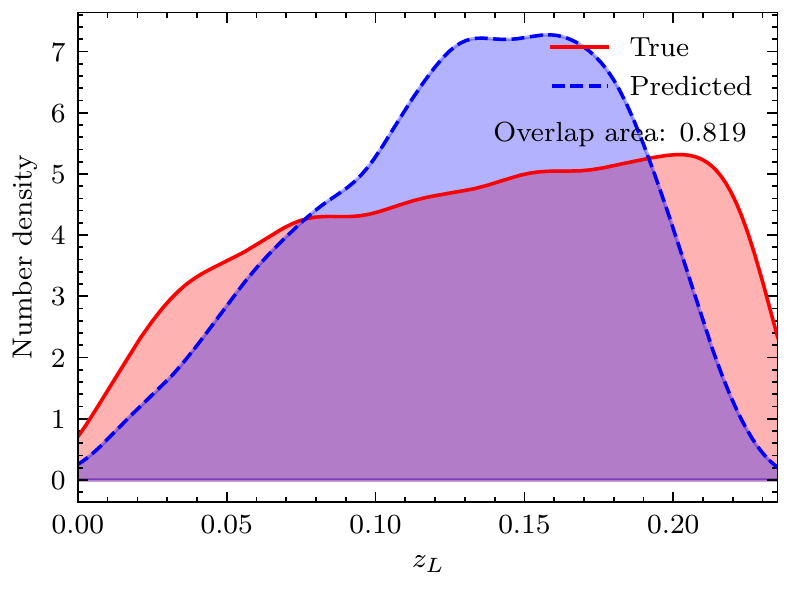}
    }
    \subfigure[$L_\mathrm{PM}$ - $y$]
    {
        \includegraphics[width=.4\linewidth]{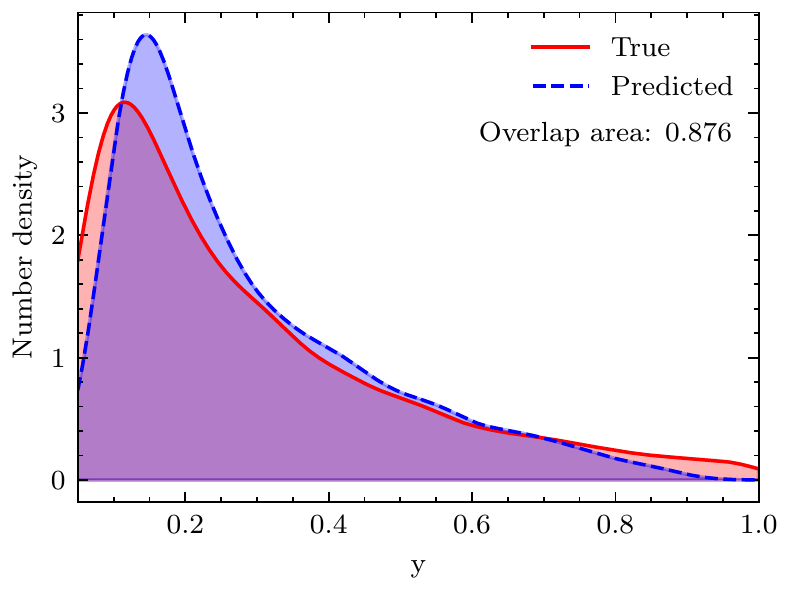}
    }
    \caption{Number density distributions of the true (red solid line) and predicted (blue dashed line) parameters, $M_\textrm{S}^c$, $M_\textrm{L}$, $z_\textrm{S}$, $z_\textrm{L}$, and $y$ of $L_\mathrm{PM}$. We can see the results are consistent to the residual plots: the distributions of predicted mass parameters follow the corresponding distribution of true parameters more or less identically. The overestimated and underestimated predictions on the lower and higher values of $z$ and $y$ parameters shown in Figure~\ref{fig:reg_residual_pm} contribute to the more populations on the peak area of predicted parameters. \label{fig:reg_overlap_pm}}
\end{figure*}


\section{Results}
\label{sec:results}

\subsection{Performance Test of Regression}
\label{sec:regression}

For the regression on the characteristic parameters of lensed GWs, we have trained VGG in Section~\ref{sec:training} to predict the parameter vector $\bm{\theta}$ of five selected parameters: four physical parameters, $M_{S}^{c}$, $M_{L}$, $z_{S}$, $z_{L}$, and one lensing parameter, $y$. We choose $y$ since it is an essential parameter in determining $\mu_\pm(y)$ and $\Delta t_d (y)$ and, eventually, $F(f)$. Note that we introduce the chirp mass of a source binary, $M_{S}^{c} = (m_1 m_2)^{2/5} (m_1 + m_2)^{-1/5}$, because it is a convenient parameter in describing the characteristics of the evolution of GW waveform~\citep{GW150914:2016prl}. We also consider the redshift factors, $z_S$ and $z_L$, instead of the distances, $D_S$ and $D_L$, because $\Delta t_d$ depends not only on $y$ but on $M_{Lz}$. For the conversion of the distances to the redshift factors, we suppose that the Hubble constant is $70~\mathrm{km} \cdot \mathrm{s}^{-1} \cdot \mathrm{Mpc}^{-1}$.

We examine the performance of regression on the evaluation data by comparing the predicted parameters to the true parameters via the residual plot (Figure~\ref{fig:reg_residual_pm}) and the number density distribution (Figure~\ref{fig:reg_overlap_pm}). Note that we discuss the results of $L_\mathrm{PM}$ only in this section because it is found that the results of $L_\mathrm{SIS}$ are quite similar to $L_\mathrm{PM}$. Instead, one can find the results of $L_\mathrm{SIS}$ from Appendix~\ref{apx:reg_sis}.

The first figure-of-merit of the regression performance is the residual plot, the residual of each $\theta$, $\theta_\mathrm{predicted} - \theta_\mathrm{true}$, against $\theta_\mathrm{true}$. In Figure~\ref{fig:reg_residual_pm}, we present the residual plots of $\bm{\theta}$ with the 1$\sigma$ confidence interval of each $\theta$'s residual. The color-bar showing the number density of true parameters. From this figure, we observe that most residuals to the densely populated range of the true parameters (the red and yellow points) distribute within the 1$\sigma$ confidence interval. On the other hand, the residuals to the sparsely populated range of the true parameters (blue points) tend to either overestimated prediction (positive residual values over the upper bound of 1$\sigma$) or underestimated prediction (negative residual values under the lower bound of 1$\sigma$). To reduce the overestimation and underestimation, having precise models for the physical parameters will be crucial: if we have such models, we can tune the VGG's training further based on the models and, eventually, will mitigate those wrong predictions.

As a second figure-of-merit, in Figure~\ref{fig:reg_overlap_pm}, we present the number density distribution of the true and predict parameters. We use the kernel density estimation of \texttt{scipy} for obtaining the curves of true $\theta$s (blue solid) and predicted $\theta$s (red dashed). For the quantitative measure of Figure~\ref{fig:reg_overlap_pm}, we compute two metrics between the distributions of the true and predicted values for each parameter: the match, $\mathcal{M}$, by integrating the overlapped area, and the Kullback-Leibler divergence, $\mathcal{D}_\mathrm{KL}$, by its definition
\begin{equation}
\mathcal{D}_\mathrm{KL} := \sum_{\theta} P_p(\theta)\log{\left( \frac{P_p(\theta)}{P_t(\theta)} \right)},
\end{equation}
where $P_t(\theta)$ and $P_p(\theta)$ denote the probability distribution of a true $\theta$ and its corresponding predicted $\theta$, respectively.
We tabulate the computed $\mathcal{M}$ and $\mathcal{D}_\mathrm{KL}$ in Table~\ref{tab:reg_metrics}. Note that if the two distributions are identical, $\mathcal{M}$ becomes 1 and $\mathcal{D}_\mathrm{KL}$ becomes $0$.

\begin{table}[t]
\caption{The match, $\mathcal{M}$, and Kullback-Leibler divergence, $\mathcal{D}_\mathrm{KL}$, between $P_t(\theta)$ and $P_p(\theta)$ of $L_\mathrm{PM}$ depicted in Figure~\ref{fig:reg_overlap_pm}. The boldfaced values show the better result in the comparison between the source and lens systems. $\mathcal{M}$ and $\mathcal{D}_\mathrm{KL}$ show consistent results that the predictions on $\theta$s of the source system are better than those of the lens system. \label{tab:reg_metrics}}
\centering
\begin{tabular}{l | c c c c | c}
\hline
\hline
Metrics & $M_S^c$ & $M_L$ & $z_S$ & $z_L$ & $y$ \\
\hline
$\mathcal{M}$ & \textbf{0.979} & 0.944 & \textbf{0.909} & 0.819 & 0.876\\
$\mathcal{D}_\mathrm{KL}$ & \textbf{0.011} & 0.021 & \textbf{0.064} & 0.152 & 0.051\\
\hline
\hline
\end{tabular}
\end{table}

\begin{figure*}[t]
    \centering
    \subfigure[Case I - PM]
    {
        \includegraphics[width=.32\linewidth]{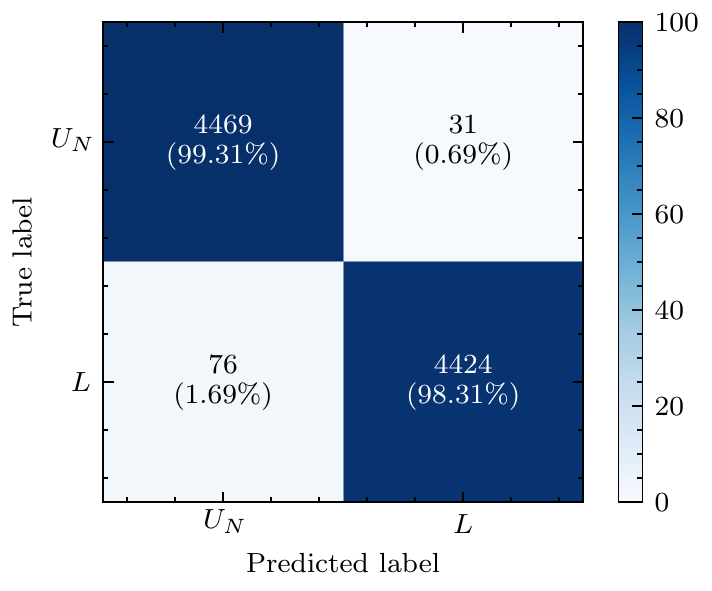}
        \label{fig:clf_cm_1_pm}
    }
    \subfigure[Case II - PM]
    {
        \includegraphics[width=.32\linewidth]{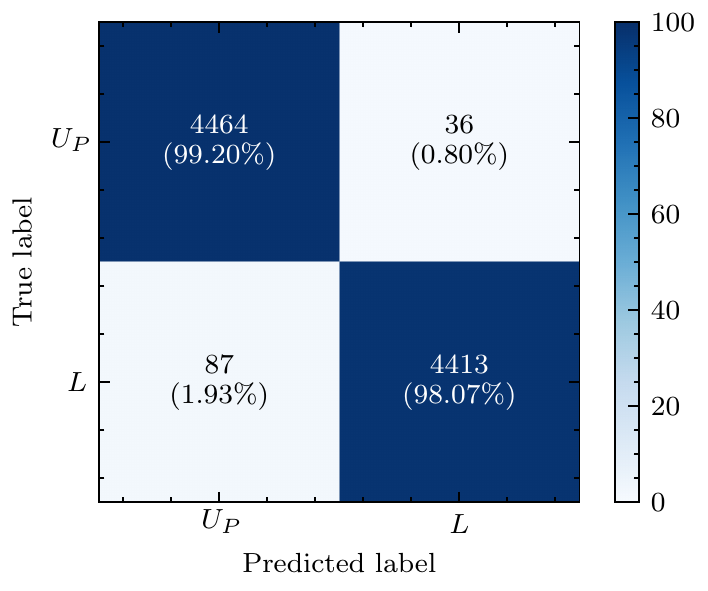}
        \label{fig:clf_cm_2_pm}
    }
    \subfigure[Case III - PM]
    {
        \includegraphics[width=.32\linewidth]{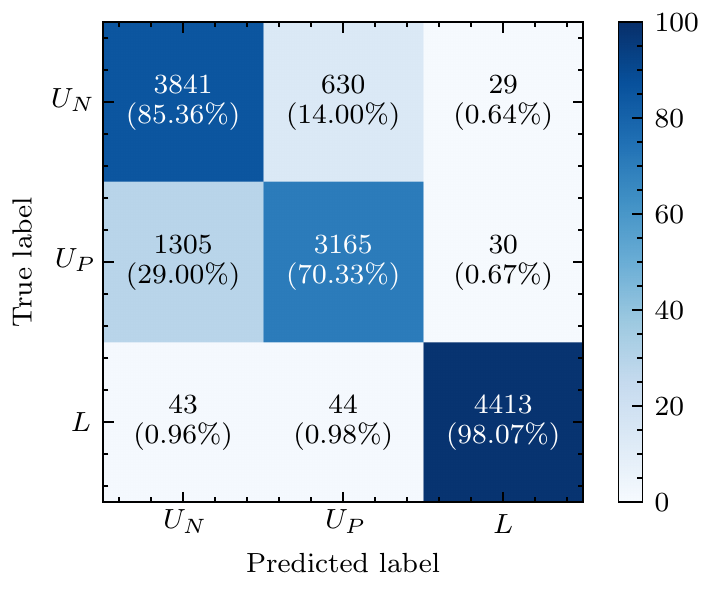}
        \label{fig:clf_cm_3_pm}
    }\\
    \subfigure[Case I - SIS]
    {
        \includegraphics[width=.32\linewidth]{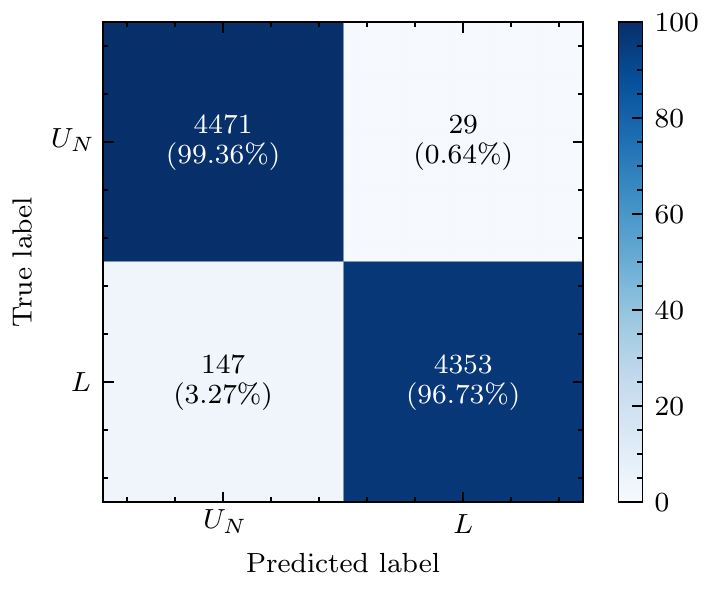}
        \label{fig:clf_cm_1_sis}
    }
    \subfigure[Case II - SIS]
    {
        \includegraphics[width=.32\linewidth]{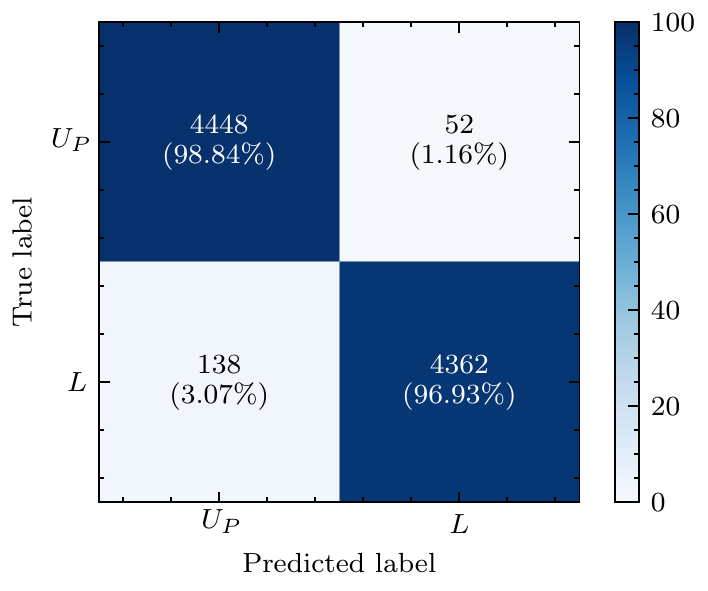}
        \label{fig:clf_cm_2_sis}
    }
    \subfigure[Case III - SIS]
    {
        \includegraphics[width=.32\linewidth]{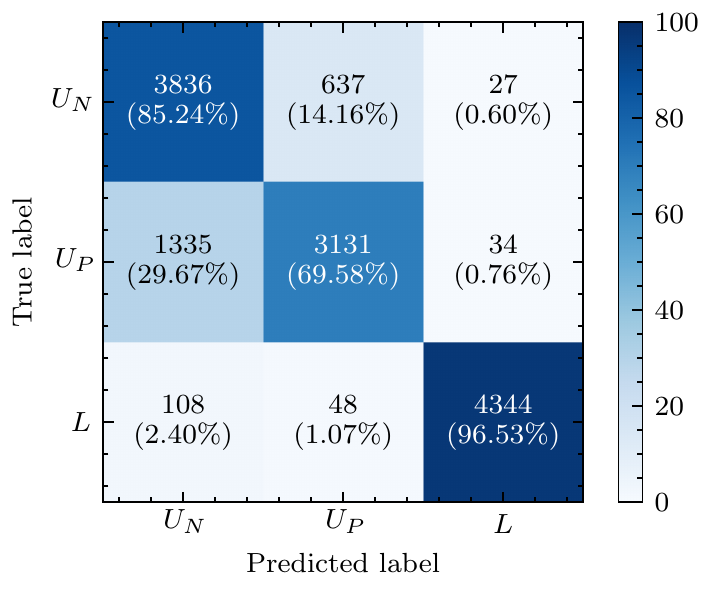}
        \label{fig:clf_cm_3_sis}
    }
    \caption{Confusion matrices for the classification on $U_N$ vs. $L$ (Case I), $U_P$ vs. $L$ (Case II), and $U_N$ vs. $U_P$ vs. $L$ (Case III) from left to right. The color bar indicates the ratio of correctly or incorrectly classified samples to the total number of samples of each signal type. The diagonal and off-diagonal cells, respectively, show correctly classified and misclassified samples. The number of classified samples is given in the middle of each cell followed by the ratio of the samples in percentage to the total number of true samples.}
    \label{fig:clf_cm}
\end{figure*} 

We can see the results shown in Figure~\ref{fig:reg_overlap_pm} are consistent with the residual plots: the distributions of predicted mass parameters follow the corresponding distribution of true parameters more or less identically. Meanwhile, the overestimated and underestimated predictions on the redshift parameters and the position parameter shown in the residual plots of Figure~\ref{fig:reg_residual_pm} contribute to more populations on the peak area of predicted parameters over the maximum number density of true parameters. Both $\mathcal{M}$ and $\mathcal{D}_\mathrm{KL}$ tabulated in Table~\ref{tab:reg_metrics} are also consistent with Figure~\ref{fig:reg_overlap_pm}: the predictions on the mass parameters are better than the redshift parameters; the predictions on the parameters of the source system are better than the lens system.

We also present the result of an additional experiment on the regression of another lensing parameter, the magnification factors, $\mu_\pm$ in Appendix~\ref{apx:reg_magnification}. The purpose of the experiment is to validate whether the regression of VGG can predict $\mu_\pm$ correctly too and it agrees with the $y$-dependency in $\mu_\pm$ or not; from the visual inspection of Figure~\ref{fig:reg_lensing_params}, the result agrees with the expected relations between the two parameters given in Equations~\eqref{eq:mu_pm} and \eqref{eq:mu_sis}.

As noted earlier, we forecast that precise models on the physical parameters of source and lens systems can enhance the identification performance on the lensing signature from the spectrograms of GWs. We expect that future observations of both the GW detectors and the upcoming EM survey telescopes will bring us more information about the source and lens systems. Then, it will help us to build precise models on the physical parameters of those systems. We will discuss the detailed prospects of using the observational results obtainable from the EM survey telescopes in Section~\ref{sec:discussion}.

On the other hand, we have sampled while investigating more appropriate black hole and gravitational lens populations can be a worthwhile long-term endeavor, we would need much more complex lensed waveforms to perform rigorous population studies, including wave optics effects, macro models, and micro/millilens fields.

\subsection{Performance Test of Classification}
\label{sec:classification}

\begin{figure*}[t!]
    \centering
    \subfigure[Case I - $U_N$ (correct)]
    {
	\includegraphics[width=.45\linewidth]{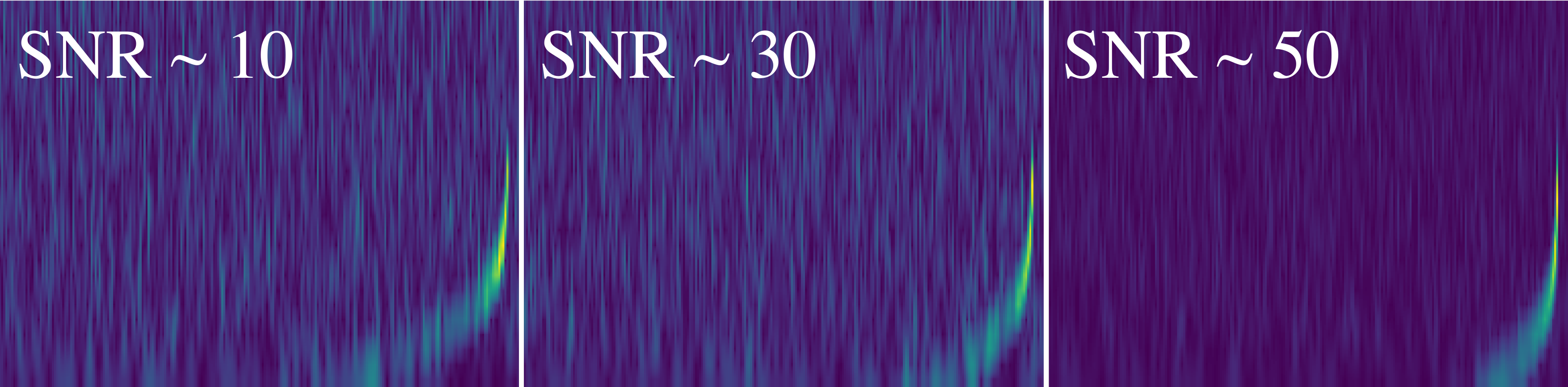}
    }
    \subfigure[Case I - $U_N$ (incorrect)]
    {
        \includegraphics[width=.45\linewidth]{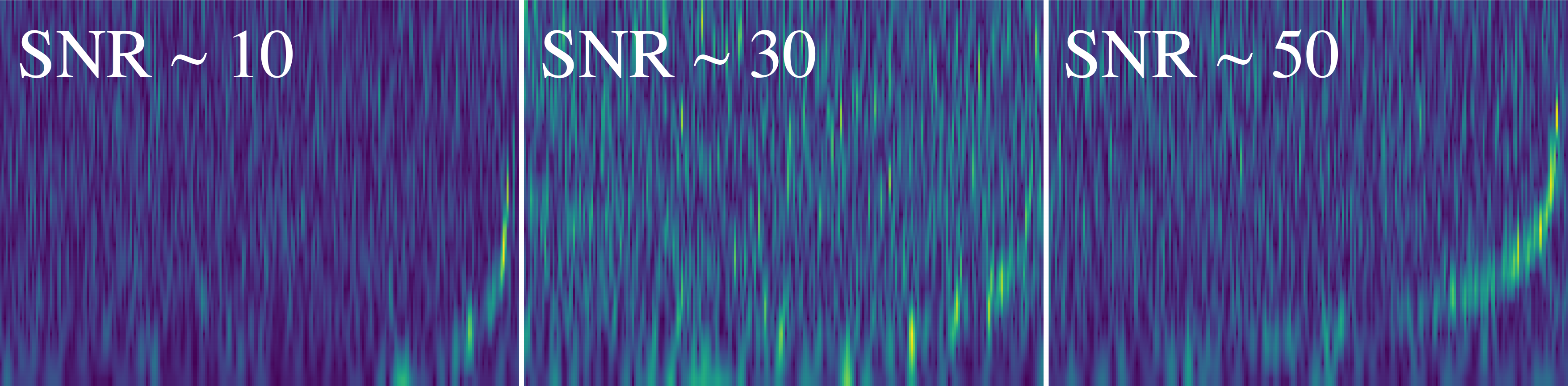}
    }\\
    \subfigure[Case I - $L_\mathrm{PM}$ (correct)]
    {
        \includegraphics[width=.45\linewidth]{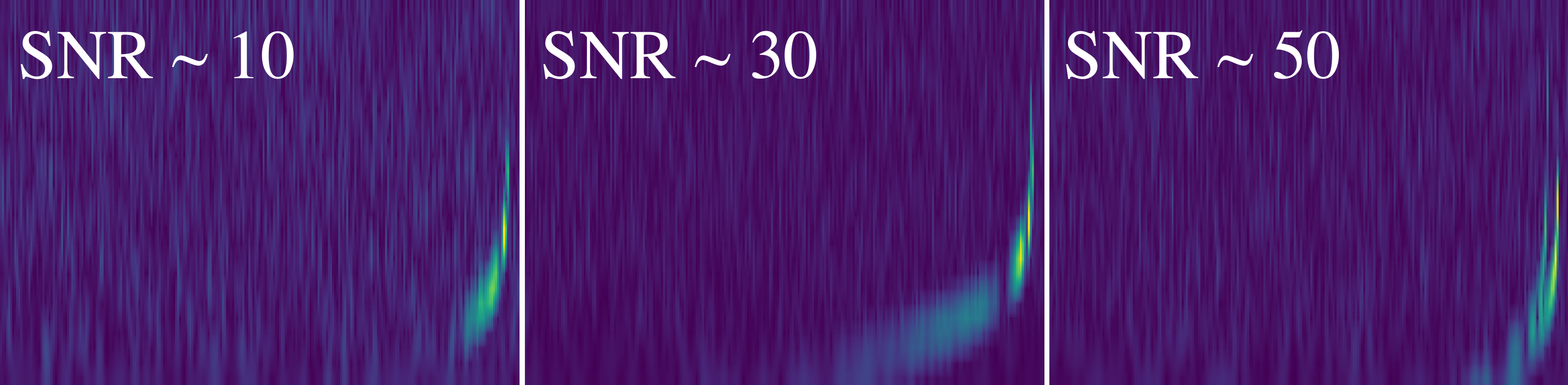}
    }
    \subfigure[Case I - $L_\mathrm{PM}$ (incorrect)]
    {
        \includegraphics[width=.45\linewidth]{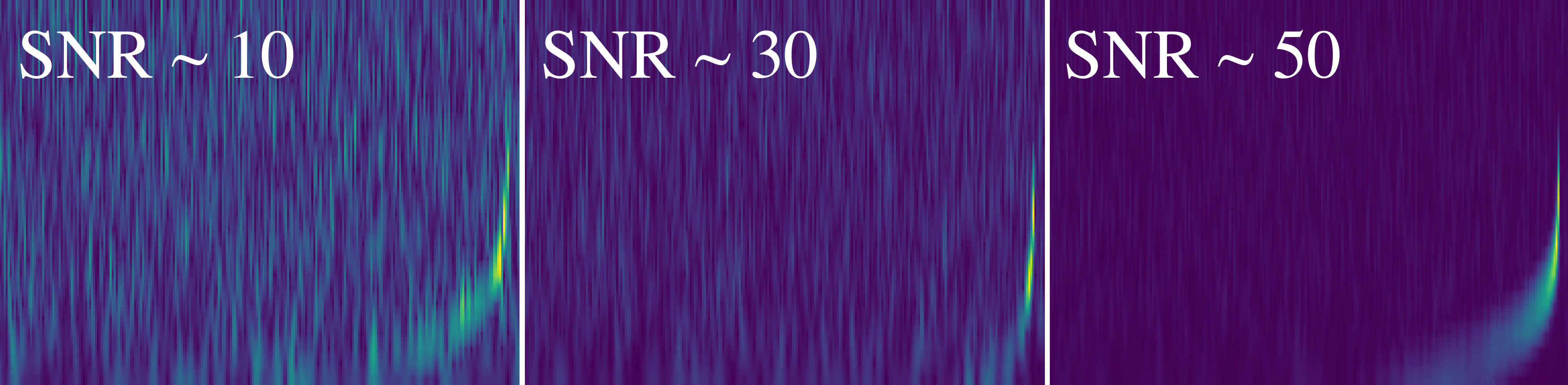}
    }\\
    \subfigure[Case I - $L_\mathrm{SIS}$ (correct)]
    {
        \includegraphics[width=.45\linewidth]{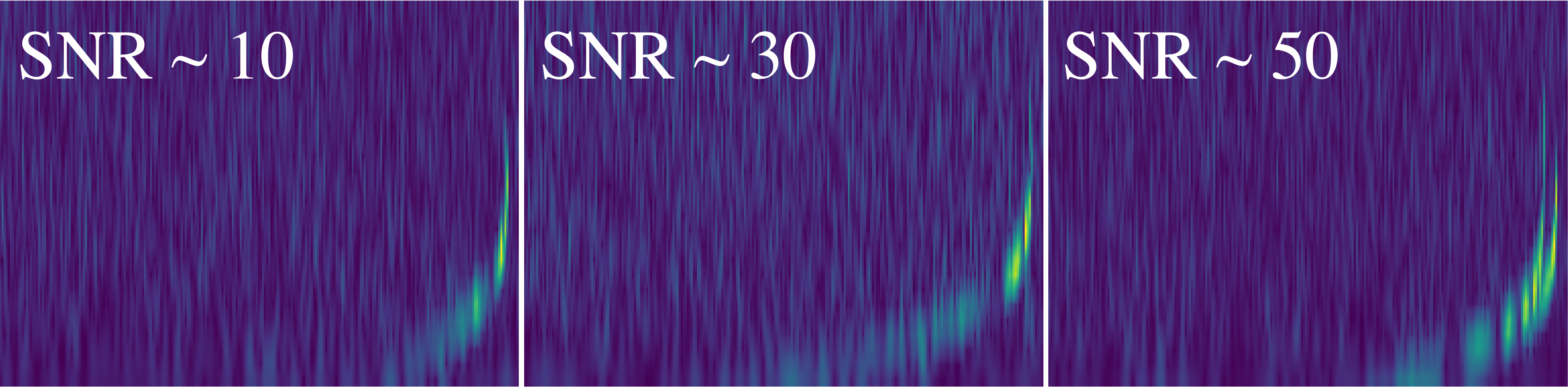}
    }
    \subfigure[Case I - $L_\mathrm{SIS}$ (incorrect)]
    {
        \includegraphics[width=.45\linewidth]{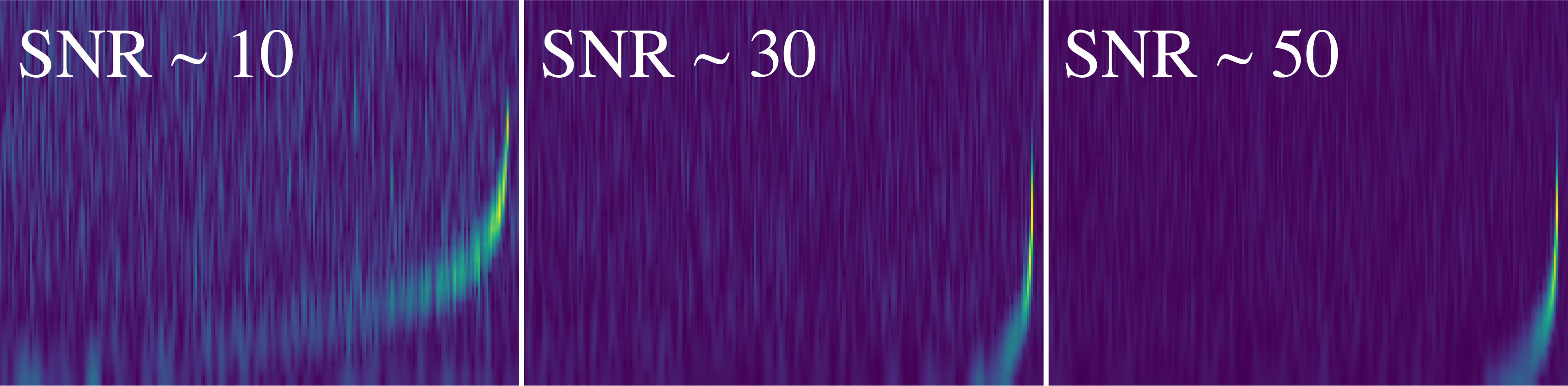}
    }\\
    \caption{Example spectrograms of correctly (left) and incorrectly (right) classified samples for Case I. Each subfigure consists of three spectrograms of SNR $\sim 10, 30, 50$ from left to right. From the right panels, we can see that the misclassified spectrograms of $L_\mathrm{PM}$ and $L_\mathrm{SIS}$ are hardly distinguishable from $U_N$ even for SNR $\gtrsim 30$. Meanwhile, the misclassified $U_N$s show false beating patterns caused by the noise. \label{fig:corr_incorr_samples_case1}}
\end{figure*}
\begin{figure*}[t!]
    \centering
    \subfigure[Case II - $U_P$ (correct)]
    {
	\includegraphics[width=.45\linewidth]{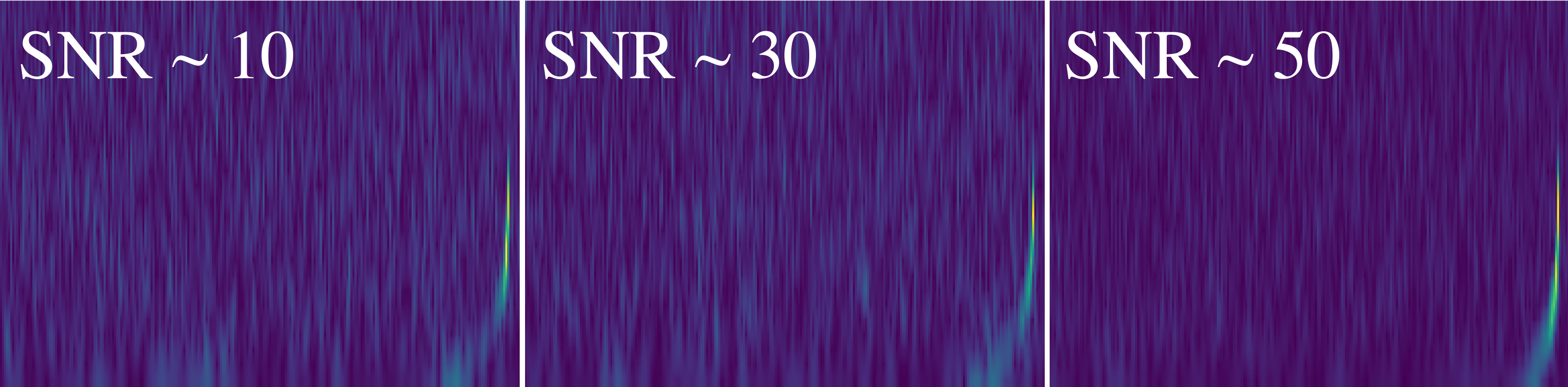}
    }
    \subfigure[Case II - $U_P$ (incorrect)]
    {
        \includegraphics[width=.45\linewidth]{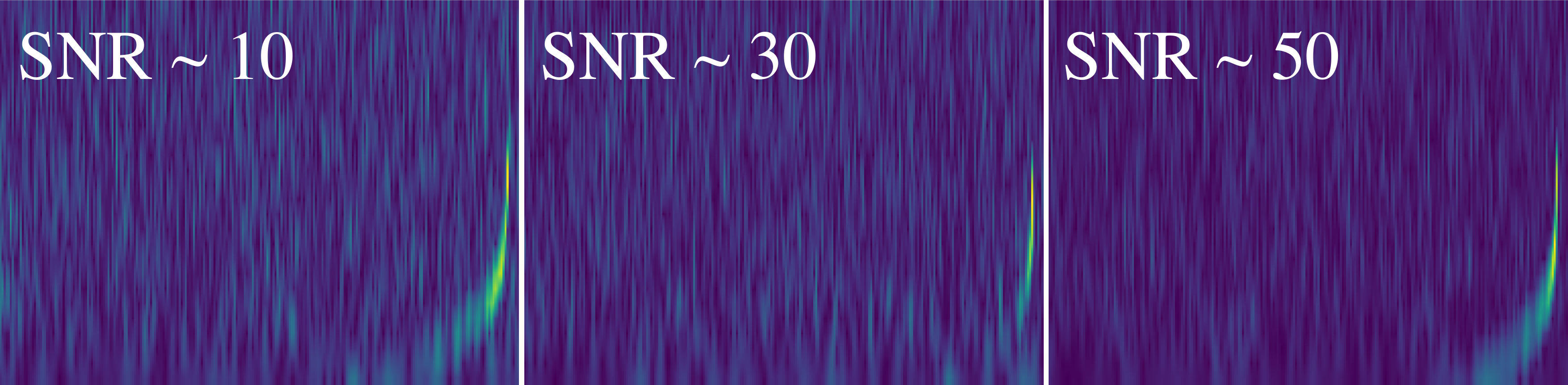}
    }\\
    \subfigure[Case II - $L_\mathrm{PM}$ (correct)]
    {
        \includegraphics[width=.45\linewidth]{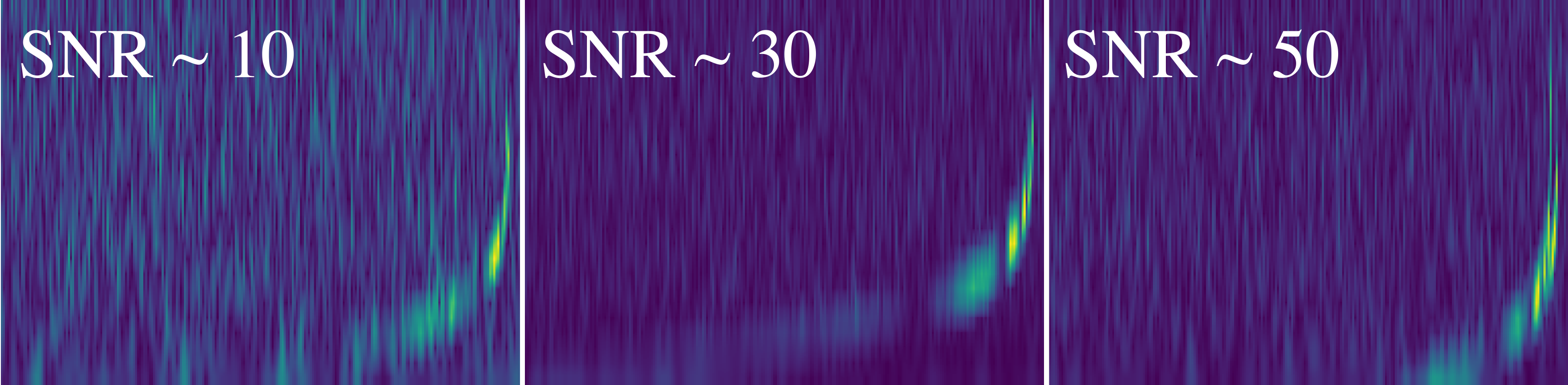}
    }
    \subfigure[Case II - $L_\mathrm{PM}$ (incorrect)]
    {
        \includegraphics[width=.45\linewidth]{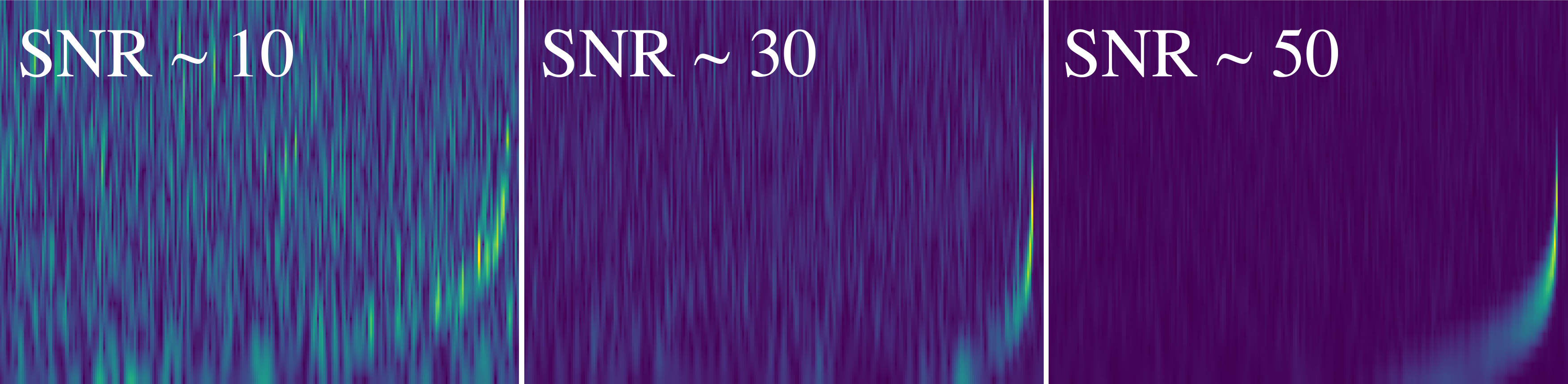}
    }\\
    \subfigure[Case II - $L_\mathrm{SIS}$ (correct)]
    {
        \includegraphics[width=.45\linewidth]{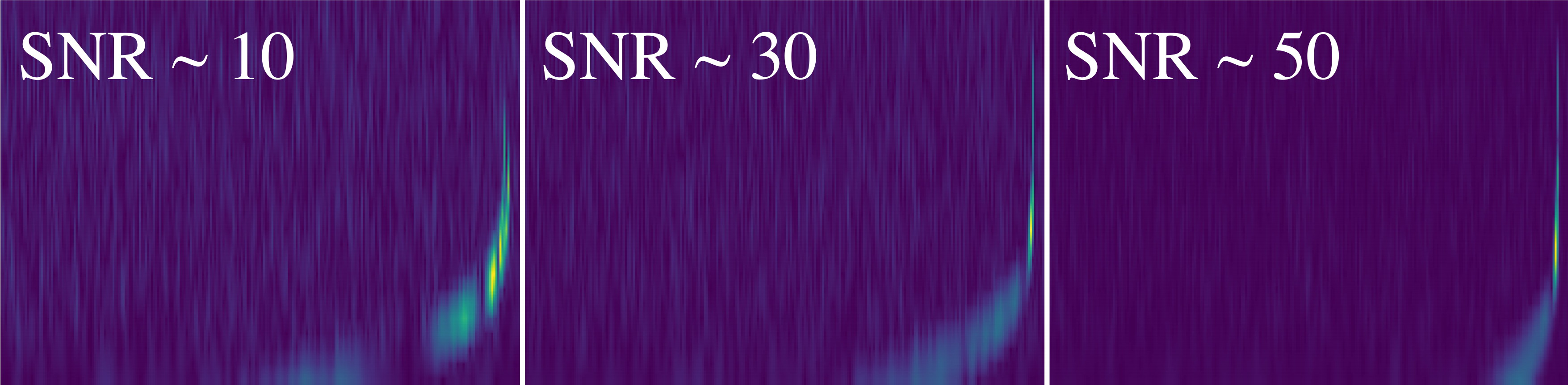}
    }
    \subfigure[Case II - $L_\mathrm{SIS}$ (incorrect)]
    {
        \includegraphics[width=.45\linewidth]{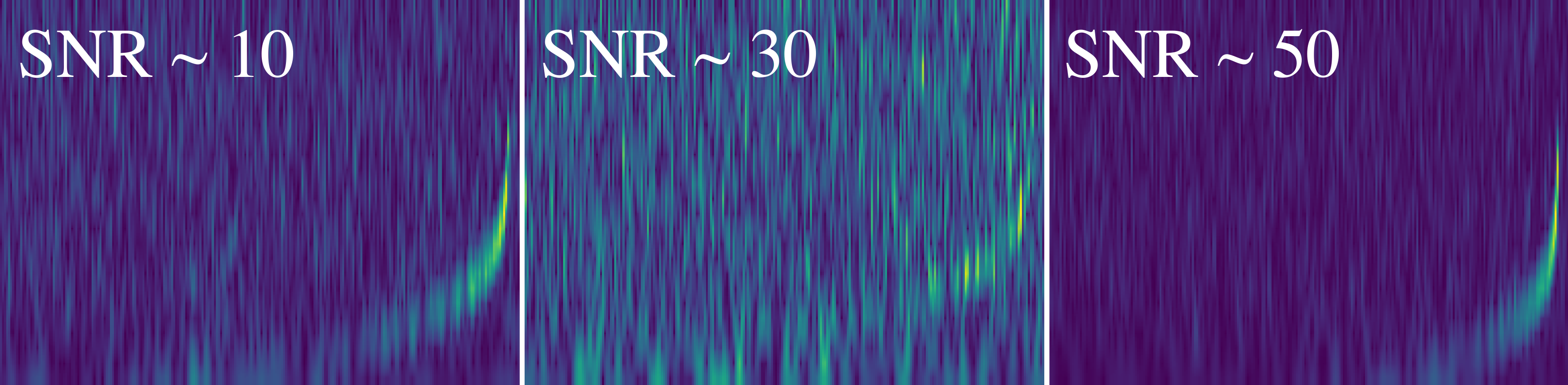}
    }\\
    \caption{The same example spectrograms of correctly (left) and incorrectly (right) classified samples for Case II. Similar to Case I of Figure~\ref{fig:corr_incorr_samples_case1}, we can see from the right panels that the spectrograms of $L_\mathrm{PM}$ and $L_\mathrm{SIS}$ are misclassified when they are hardly distinguishable from $U_P$ or show a rather noisy spectrogram, although its SNR is $\sim$ 30. \label{fig:corr_incorr_samples_case2}}
\end{figure*}

For the evaluation of classification performance, we firstly investigate whether the trained VGG correctly distinguishes the spectrograms of $L$ from $U_N$ (Case I) or $U_P$ (Case II). Next, we document our ability to discriminate between all three different types, that is, $L$, $U_N$, and $U_P$ (Case III). 
We separate $L$ further into $L_\mathrm{PM}$ and $L_\mathrm{SIS}$ depending on the lens model. We calculate the confusion matrix for a figure-of-merit of classification performance and present it in Figure~\ref{fig:clf_cm} with the number and ratio of correctly and incorrectly classified samples.

\begin{figure*}[t!]
    \centering
    \subfigure[Case I - $L_\mathrm{PM}$]
    {
        \includegraphics[width=.4\linewidth]{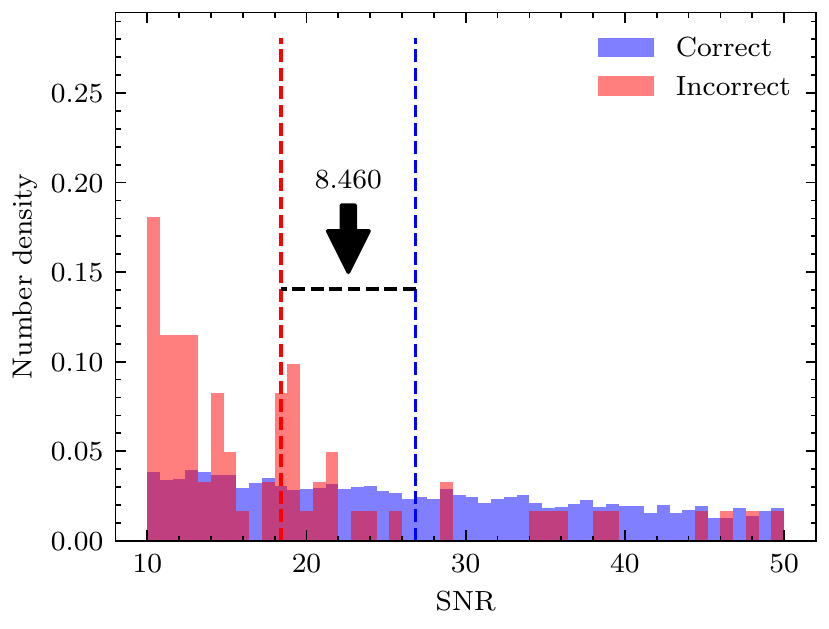}
    }
    \subfigure[Case I - $L_\mathrm{SIS}$]
    {
        \includegraphics[width=.4\linewidth]{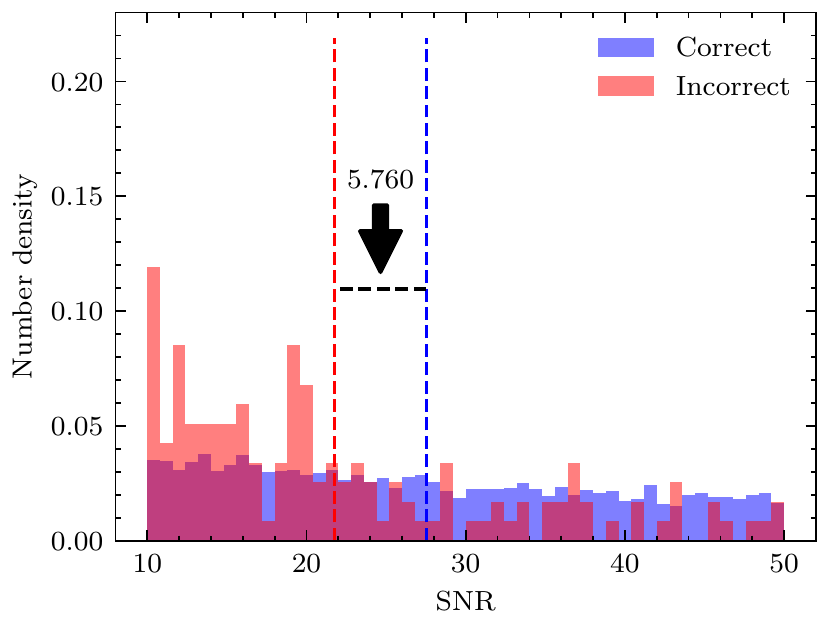}
    }\\
    \subfigure[Case I - $L_\mathrm{PM}$ (SNR $\sim$30)]
    {
        \includegraphics[width=.4\linewidth]{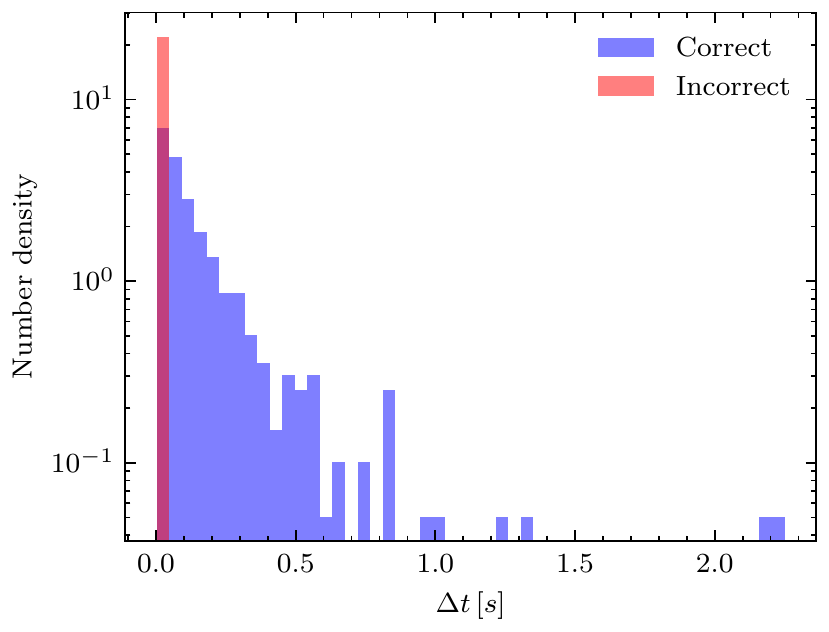}
    }
    \subfigure[Case I - $L_\mathrm{SIS}$ (SNR $\sim$30)]
    {
        \includegraphics[width=.4\linewidth]{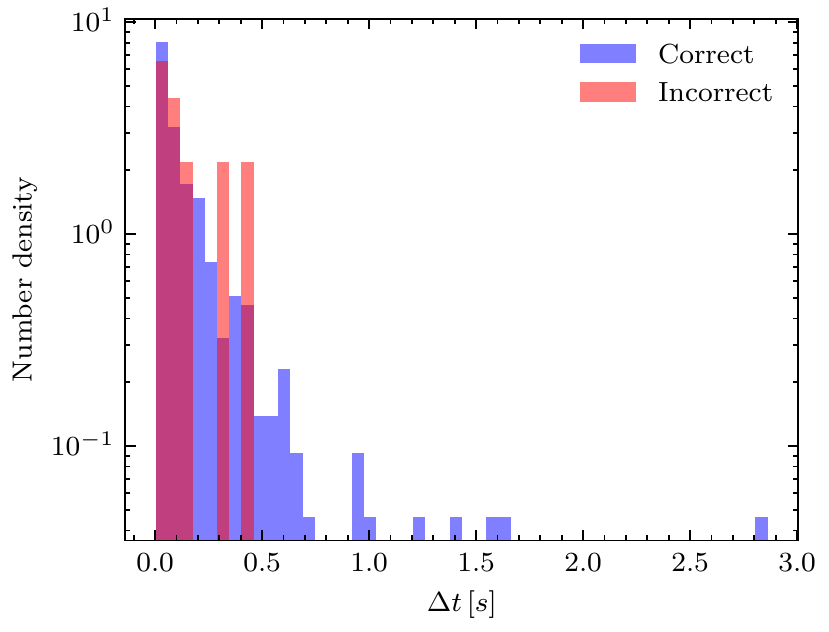}
    }
    \caption{Distributions of the number density of $L$ samples in Case I with respect to the SNR (top) and $\Delta t_d$ corresponding to SNR $\sim$30 (bottom). The blue and red bars represent the number density of correctly classified samples and incorrectly classified ones, respectively. The annotated number above the black arrow in the top panels shows the value of $\Delta$$\left< \mathrm{SNR} \right>$. We observe that the bigger $\Delta$$\left< \mathrm{SNR} \right>$ of $L_\mathrm{PM}$ than $L_\mathrm{SIS}$ is consistent with the higher ratio of correctly classified $L_\mathrm{PM}$ samples than $L_\mathrm{SIS}$ samples shown in Figure~\ref{fig:clf_cm}. On the other hand, we find that most misclassified $L$ samples correspond to $\Delta t_d < 0.25$ sec.\label{fig:pdf}}
\end{figure*}

The trained VGGs of both Case I and II can correctly classify expected types over 96\% of the evaluation data as shown in the diagonal cells of the left and middle panels of Figure~\ref{fig:clf_cm}. When we look at the spectrogram examples of correctly classified and incorrectly classified samples (Figures~\ref{fig:corr_incorr_samples_case1} and \ref{fig:corr_incorr_samples_case2}), we can see that correctly classified $L$ samples, both $L_\mathrm{PM}$ and $L_\mathrm{SIS}$, show recognizable beating patterns distinguishable from $U_N$ or $U_P$. However, the beating pattern of incorrectly classified $L_\mathrm{PM}$ and $L_\mathrm{SIS}$ samples are not easily distinguishable from $U_N$ or $U_P$ even for higher SNRs $\gtrsim 30$. 

In the comparison between $L_\mathrm{PM}$ and $L_\mathrm{SIS}$ from the confusion matrix (see the bottom-right cells in Figure~\ref{fig:clf_cm}), we can see that $\sim 98\%$ of $L_\mathrm{PM}$ samples are correctly classified while $\sim$ 97\% of $L_\mathrm{SIS}$ samples are correctly classified. To understand the small difference, we examine the distributions of the number density of $L_\mathrm{PM}$ and $L_\mathrm{SIS}$ of Case~I for SNR (top panels of Figure~\ref{fig:pdf}). From the figure, we observe that the distribution looks consistent with the confusion matrix and incorrectly classified samples of both $L_\mathrm{PM}$ and $L_\mathrm{SIS}$ mostly distributed in the lower SNR region, SNR $\lesssim$ 20 as expected. However, for $L_\mathrm{SIS}$, there are more incorrectly classified samples than $L_\mathrm{PM}$, even for SNR $\gtrsim$ 30.

We also calculate the mean SNRs of correctly/incorrectly classified samples and compute the difference between them with denoting the difference as $\Delta$$\left< \mathrm{SNR} \right>$; it turns out that $\Delta$$\left< \mathrm{SNR} \right>_\mathrm{PM}$ $>$ $\Delta$$\left< \mathrm{SNR} \right>_\mathrm{SIS}$. It implies that the beating pattern of $L_\mathrm{SIS}$ samples might be slightly less discernible from $U_N$ or $U_P$ than $L_\mathrm{PM}$ samples, even with higher SNRs. From this observation, we recognize that the classification performance of the VGG is certainly limited when the beating pattern is insignificant. We also find the requiring higher SNR is not the only condition for the clear identification of the lensing effect.

Further, for those incorrectly classified $L$ samples, we draw the distribution of number density with respect to $\Delta t_d$ of the $L$ samples corresponding to SNR $\sim 30$ (bottom panels of Figure~\ref{fig:pdf}) and find that the $L$ samples generated with the parameters resulting in $\Delta t_d < 0.25$ sec are mostly misclassified. Therefore, we find that the trained VGG has another limit in identifying lensed GW signals formed by too short $\Delta t_d$.

When we compare the confusion matrices of Case III (right panels in Figure~\ref{fig:clf_cm}) to those of Cases I and II, we find that the ratios of correctly classified $L$ samples are similar to or slightly smaller than other cases. In specific, the proportion of correctly classified samples significantly decreased by $\sim$14\% for the $U_N$ and $\sim$28\% -- 29\% for the $U_P$. In contrast, it is reduced by only $\lesssim$ 0.4\% for $L$. Thus, we can see that the beating pattern caused by the lensed GWs is still sufficiently significant than the unlensed ones, either $U_N$ or $U_P$.

In the meantime, from a comparison between $U_N$ and $U_P$ of Case III, we see that the classification of VGG on $\sim$14\% -- 29\% of samples is incorrect; it misclassifies $U_N$ as $U_P$, or vice versa. The ratio of $U_P$ to $U_N$ is $\gtrsim 15\%$ higher than the opposite case. We suspect that the effect of the precession-induced modulation might be less visible than the lensing-induced beating patterns. However, we leave enhancing the discrepancy between $U_N$ and $U_P$ to future work because it is out of interest of this study.

\subsection{Validation Test}
\label{sec:validation}

In this section, we examine the validity of our method to the population properties of the GW signals' progenitors of GWTC-2. First, we select the population models of the progenitor system from~\cite{Abbott:2020gyp}: the `Power law + Peak' model for $m_1$ and $m_2$; the `Non-Evolving' model for $z_S$. Here, we consider the non-spinning binary black hole system only for convenience. On the other hand, for the lens system, we take the distribution of $M_L$ to be the same as that of $M_L$ in this work and let $z_L$ be half of the value of the sampled $z_S$ for simplicity. With the populations, we obtain $\sim 300$ lensed samples for both $L_\mathrm{PM}$ and $L_\mathrm{SIS}$ satisfying $12 <  \mathrm{SNR} \leq 13$. Next, we evaluate the validation data with the VGGs of PM and SIS lens models (VGG-PM and VGG-SIS, respectively) trained for either the regression or the Case I classification. 

We summarize the resulting $\mathcal{D}_\mathrm{KL}$s in Table~\ref{tab:reg_validation} with the $\mathcal{D}_\mathrm{KL}$s computed in Sec.~\ref{sec:regression} or Appendix \ref{apx:reg_sis} with our evaluation data for comparison. One can see that $\mathcal{D}_\mathrm{KL}$s of the validation data of both $L_\mathrm{PM}$ and $L_\mathrm{SIS}$ are higher than $\mathcal{D}_\mathrm{KL}$s of our evaluation data while $\mathcal{D}_\mathrm{KL}$s of the validation data also show better results on $M_S$ than $M_L$ and $z_S$ than $z_L$, consistently. Meanwhile, the accuracy of the classification of the validation data obtained from VGG-PM and VGG-SIS are, respectively, $\sim$68\% for $L_\mathrm{PM}$ and $\sim$62\% for $L_\mathrm{SIS}$.

\begin{table}[t]
\caption{$\mathcal{D}_\mathrm{KL}$s between $P_t(\theta)$ and $P_p(\theta)$ of the selected five parameters of $L_\mathrm{PM}$ or $L_\mathrm{SIS}$ of the validation data. For comparison, we also tabulate $\mathcal{D}_\mathrm{KL}$s summarized in Tables~\ref{tab:reg_metrics} and \ref{tab:reg_metrics_sis}. The bold-faced values indicate better results in the comparison between the source system and lens system. On the other hand, we mark the asterisk on better results among the validation data and our evaluation data. \label{tab:reg_validation}}
\centering
\begin{tabular}{l | c c c c | c}
\hline
\hline
Data & $M_S^c$ & $M_L$ & $z_S$ & $z_L$ & $y$ \\
\hline
$L_\mathrm{PM}^\mathrm{val}$ & \textbf{0.140} & 0.121 & \textbf{0.381} & 0.314 & 1.318\\
$L_\mathrm{PM}^\mathrm{ours}$ & $\textbf{0.011}^*$ & $0.021^*$ & $\textbf{0.064}^*$ & $0.152^*$ & $0.051^*$\\
\hline
$L_\mathrm{SIS}^\mathrm{val}$ & \textbf{0.260} & 0.088 & \textbf{0.400} & 0.418 & 1.542\\
$L_\mathrm{SIS}^\mathrm{ours}$ & $\textbf{0.0205}^*$ & $0.0207^*$ & $\textbf{0.051}^*$ & $0.111^*$ & $0.002^*$\\
\hline
\end{tabular}
\end{table}

It is generally well-known that a trained deep learning model is sensitive and efficient to the samples from a population resembling the parameter distribution of the data used in training~\citep{hendrycks17baseline}. In this sense, the degraded performance of the pre-trained models on the newly prepared validation data is expectable. Therefore, the results of this validation test address the necessity of the complete population models on the parameters again for the successful application of our approach to the search for weakly lensed GW signals.


\section{Discussion}
\label{sec:discussion}

In this work, we considered the range of the lens's mass as $10^3\msun$ -- $10^5 \msun$. The mass range is much smaller than the known masses of galaxies that cover from $\sim 10^9 M_\odot$ to $\sim 10^{12} M_\odot$, which can cause potentially observable strong gravitational-wave lensing in the ground-based GW detectors. Instead, the chosen mass range may correspond to the mass range of certain astrophysical compact objects such as intermediate-mass black holes~\citep{imbh} or compact dark matter~\citep{Jung:2017flg}. Thus, if we can identify lensed GWs produced by the lens systems discussed in this work, the identification may help to understand the characteristics of the lens objects.

We have discarded the lensing of precessing GWs to simplify the problem. However, as shown in the classification of Case II, precessing effect itself is mostly distinguishable from the lensed GWs while it is less distinguishable than non-precessing GWs in the classification of Case III. Therefore, further systematic study about the degeneracies between the precessional and lensing parameters should be done along with the identification.

As seen from the regression result, the uncertainty in the prediction on the characteristic parameters related to the lens was much larger than that on the parameters related to the source. In other words, we may conclude that the degenerated information contained in the spectrogram is not sufficient to characterize the lens fully. For this issue, we believe that additional information obtained from the EM observations will be helpful in reducing the uncertainty and, consequently, in enhancing the prediction power of the regression about the lens. For example, for a lensed GW, we may conduct finer estimation on the physical parameters of the lens by finding coincident lensing events in the optical band survey with the LSST\footnote{Large Synoptic Survey Telescope~\citep{Ivezic:2008fe}. Recently the name has been changed to Vera C. Rubin Observatory.}~\citep{collaboration2009lsst}. Meanwhile, for the lens system like galaxy or galaxy cluster, we may understand more details of physical properties of the lens by the near-infrared observation with the SPHEREx\footnote{Spectro-Photometer for the History of the Universe, Epoch of Reionization, and Ices Explorer~\citep{spherex}}~\citep{dor2016science}. Therefore, we expect that the identification of lensed GWs will be one of the exciting fields where the multimessenger observations will be impactful.

\begin{figure}[t]
\centering
    \subfigure
    {
        \includegraphics[width=.8\linewidth]{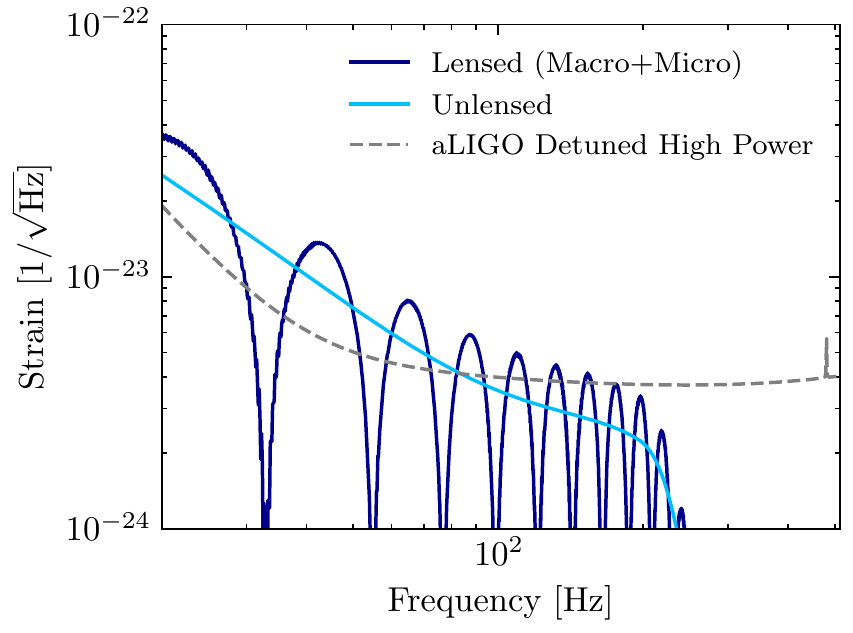}
    }\\
    \subfigure
    {
        \includegraphics[width=.8\linewidth]{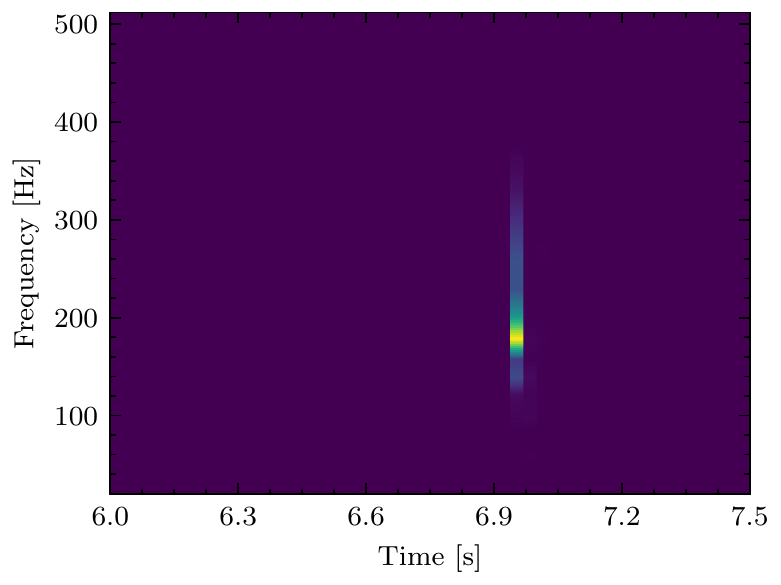}
    }
\caption{Strain of a microlensed GW (top) and its spectrogram (bottom) generated with the parameters and tools summarized in Appendix~\ref{apx:microlensing_params}.}
\label{fig:microlensing_spectrogram}
\end{figure}

On the other hand, within the sensitive frequency band of Advanced LIGO and Virgo, the possibility of observing GWs lensed by the microlenses embedded in a galaxy or galaxy cluster is discussed in~\citet{Diego:2019lcd,Pagano:2020rwj,Cheung:2020okf}. For example, it is shown that the evolution of $F(f)$ of this scenario will be non-negligible (see Figure 6 of \citet{Pagano:2020rwj}) and GWs lensed by the microlenses around such macro lens systems can be observed (see Figure 10 of \citet{Pagano:2020rwj}). Consequently, we can seamlessly anticipate seeing beating patterns in the spectrograms of such microlensed events' signals. As an example, we simply evaluate a spectrogram sample prepared by using \texttt{lensingGW} package~\citep{Pagano:2020rwj, lenstronomy}. The strain of this test and its spectrogram sample are presented in Figure~\ref{fig:microlensing_spectrogram} and the parameters used for generating this sample are summarized in Appendix~\ref{apx:microlensing_params}. We perform the classification of the microlensing samples with the VGGs used in Secs.~\ref{sec:classification} and \ref{sec:validation}. As the result, the VGG-SIS correctly classifies this sample with $100\%$ accuracy to $L$ while the VGG-PM incorrectly classifies this sample with $37\%$ accuracy to $L$. Therefore, we can ascertain from this test that there are plentiful prospects for improving the identification of the microlensing signature from spectrograms even with more complicated lens models. To do that, we can train the VGG with a new training set of data prepared by using the \texttt{lensingGW} package. However, we leave the full application of our method to the search of microlensing in GWs to future work.

\begin{figure*}[t]
    \centering
    \subfigure[$L_\mathrm{SIS}$ - $M_S^{c}$]
    {
        \includegraphics[width=.32\linewidth]{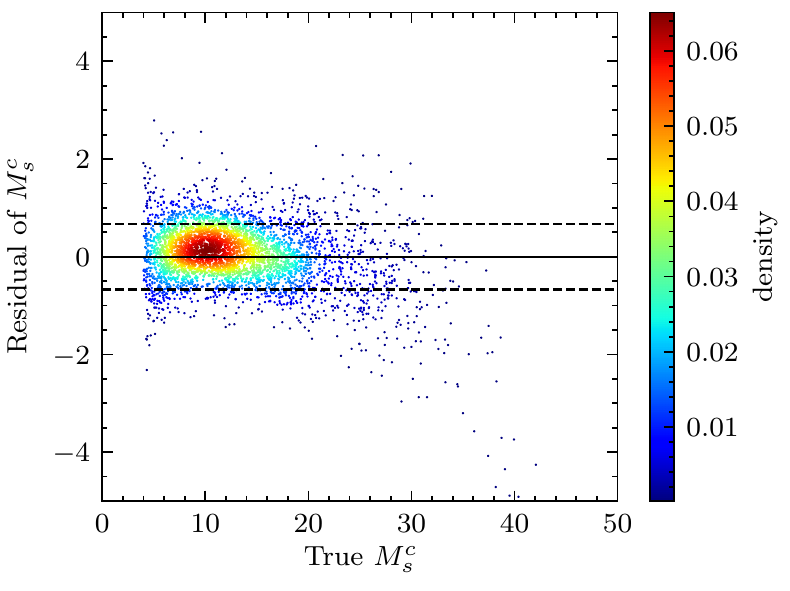}
    }
    \subfigure[$L_\mathrm{SIS}$ - $M_L$]
    {
        \includegraphics[width=.32\linewidth]{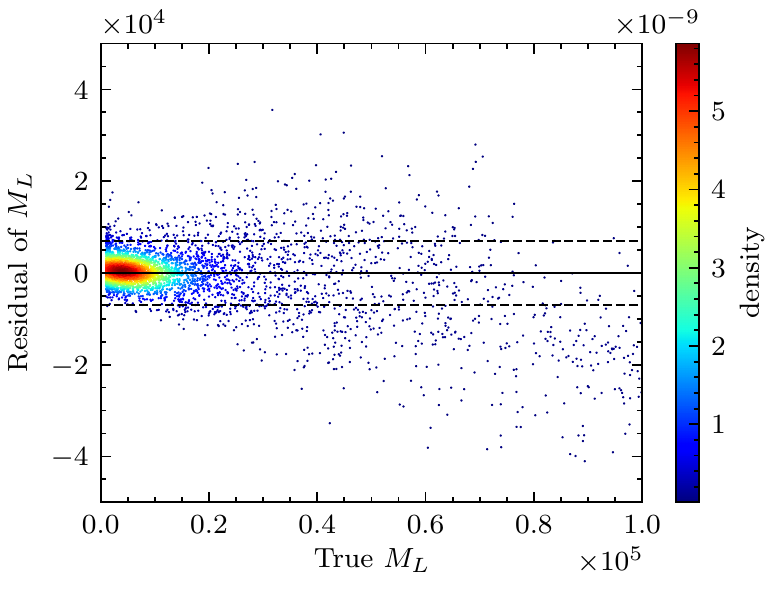}
    }\\
    \subfigure[$L_\mathrm{SIS}$ - $z_S$]
    {
        \includegraphics[width=.32\linewidth]{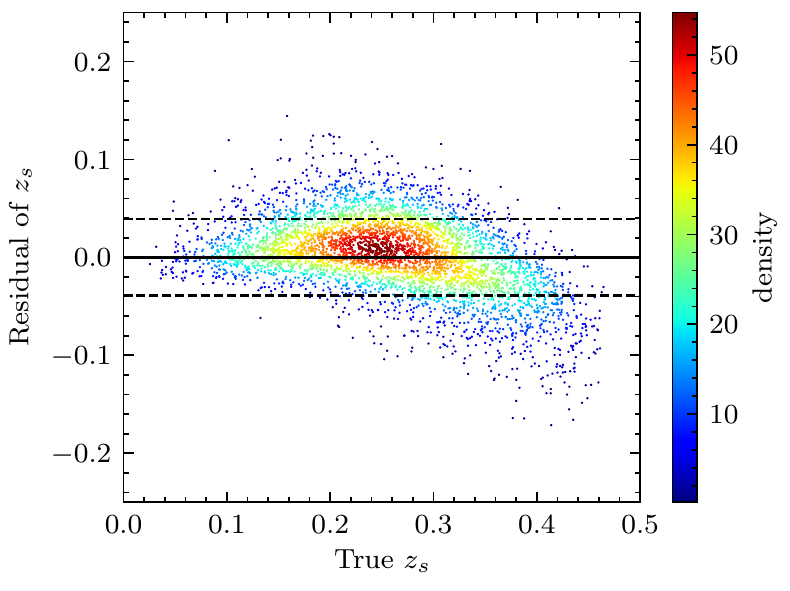}
    }
    \subfigure[$L_\mathrm{SIS}$ - $z_L$]
    {
        \includegraphics[width=.32\linewidth]{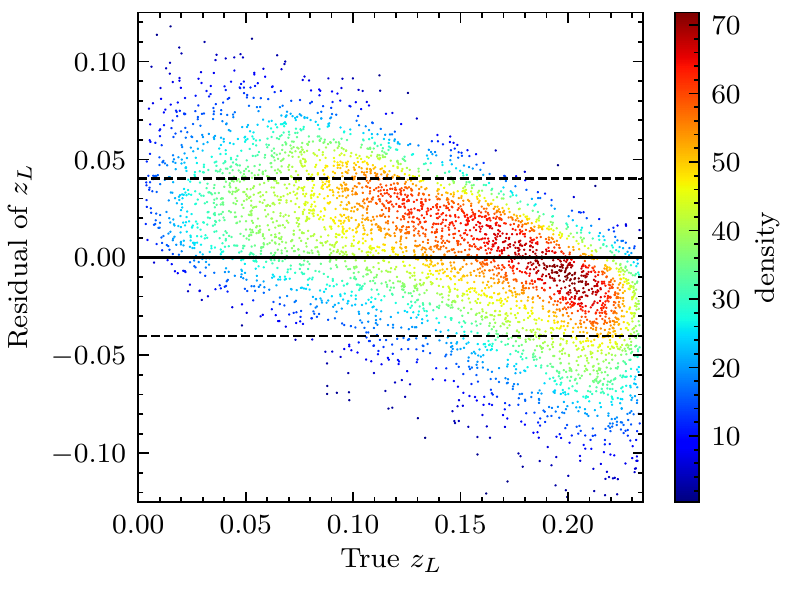}
    }
    \subfigure[$L_\mathrm{SIS}$ - $y$]
    {
        \includegraphics[width=.32\linewidth]{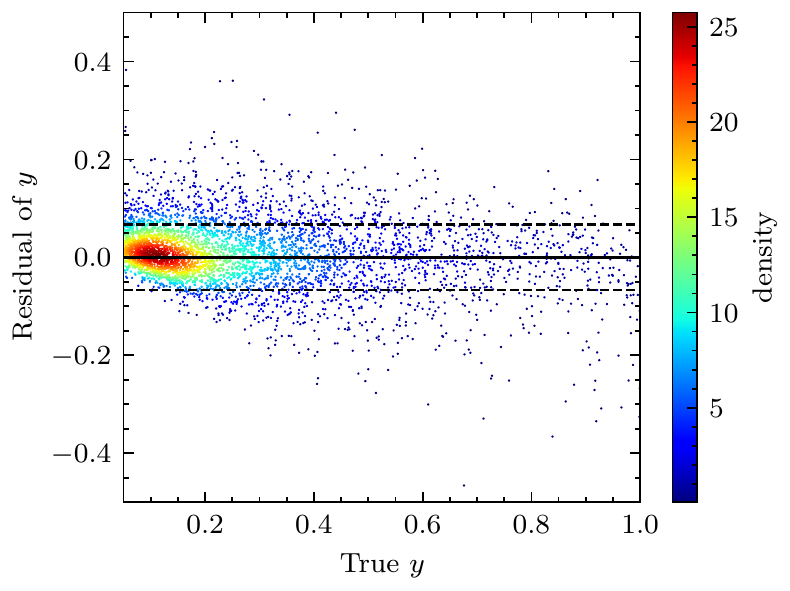}
    }
    \caption{Residual plots of $M_\textrm{S}^c$, $M_\textrm{L}$, $z_\textrm{S}$, $z_\textrm{L}$, and $y$ of $L_\mathrm{SIS}$.
    \label{fig:reg_residual_sis}}
\end{figure*}


\acknowledgments
KK thanks Min-Su Shin for constructive discussion and Giulia Pagano for useful discussion in the use of \texttt{lensingGW}. We also thank the Global Science experimental Data hub Center (GSDC) at KISTI for supporting GPU-based computing resource. KK is supported by the National Research Foundation of Korea (NRF) grant funded by the Korea government (MSIT) (NRF-2020R1C1C1005863). OAH is supported by the research program of the Netherlands Organization for Scientific Research (NWO). TGFL is partially supported by grants from the Research Grants Council of Hong Kong (Project No. 14306218), Research Committee of the Chinese University of Hong Kong, and the Croucher Foundation of Hong Kong.


\appendix

\section{Details of VGG Implementation}

\subsection{Structural Details}
\label{apx:vgg}

\begin{figure*}[t]
    \centering
    \subfigure[$L_\mathrm{SIS}$ - $M_S^{c}$]
    {
        \includegraphics[width=.32\linewidth]{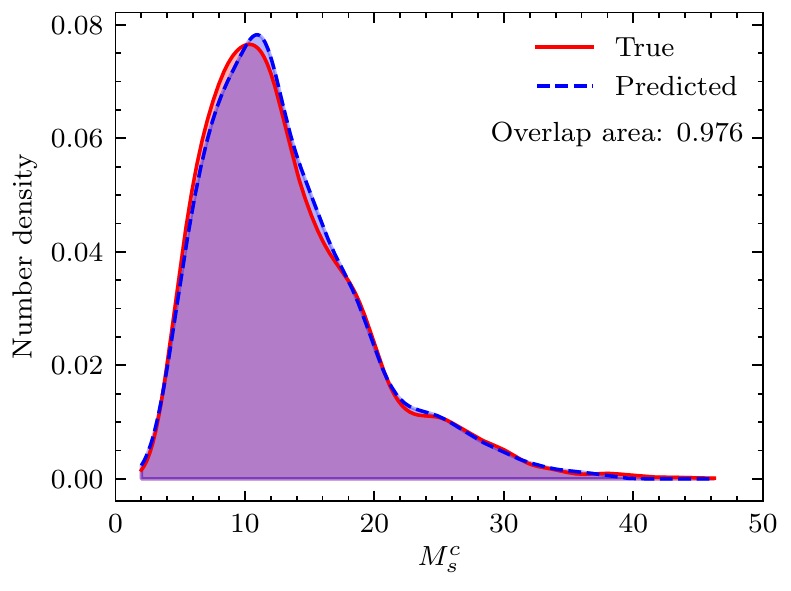}
    }
    \subfigure[$L_\mathrm{SIS}$ - $M_L$]
    {
        \includegraphics[width=.32\linewidth]{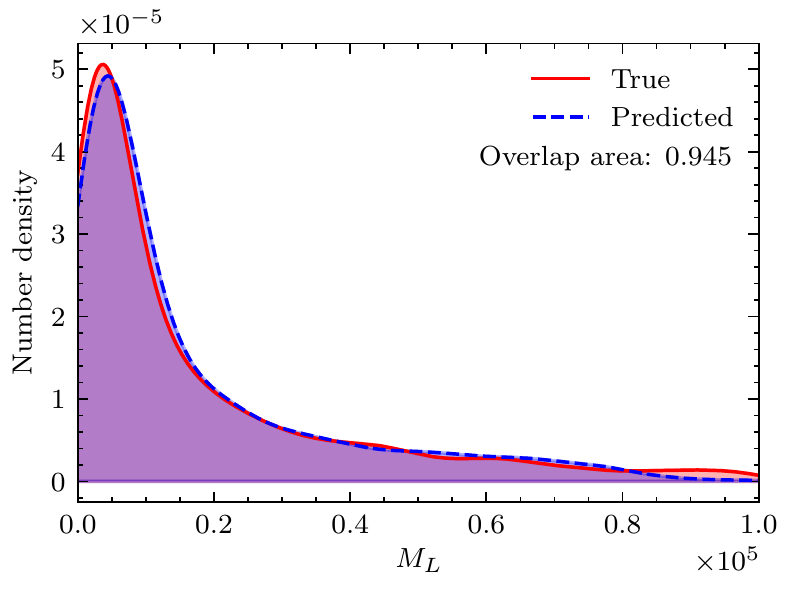}
    }\\
    \subfigure[$L_\mathrm{SIS}$ - $z_S$]
    {
        \includegraphics[width=.32\linewidth]{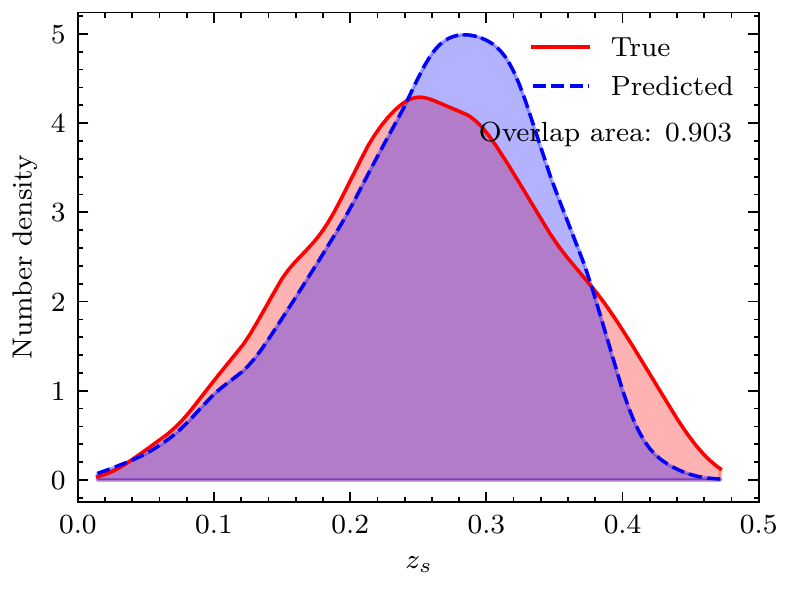}
    }
    \subfigure[$L_\mathrm{SIS}$ - $z_L$]
    {
        \includegraphics[width=.32\linewidth]{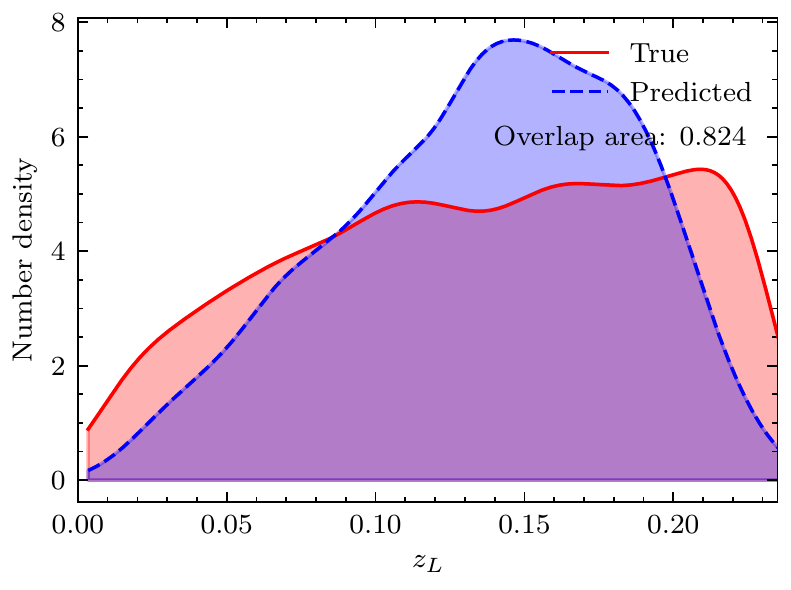}
    }
    \subfigure[$L_\mathrm{SIS}$ - $y$]
    {
        \includegraphics[width=.32\linewidth]{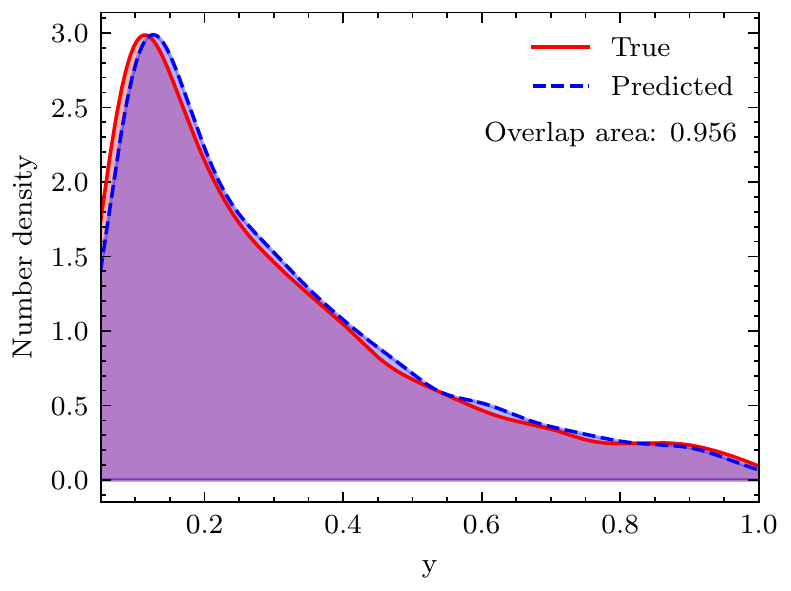}
    }
    \caption{Distributions of the true and predicted parameters of $M_\textrm{S}^c$, $M_\textrm{L}$, $z_\textrm{S}$, $z_\textrm{L}$, and $y$ of $L_\mathrm{SIS}$.
    \label{fig:reg_overlap_sis}}
\end{figure*}

We use 10 convolutional layers which consist of 3x3 convolutional filters. The number of filters in order are [16, 16, 16, 32, 32, 32, 64, 64, 64, 64, 64, 128, 128, 128, 128, 128, 128, 128, 128, 128, 128]. The strides for each convolutional layers are [1x1, 1x1, 2x2, 1x1, 1x1, 2x2, 1x1, 1x1, 1x1, 1x1, 2x2, 1x1, 1x1, 1x1, 1x1, 2x2, 1x1, 1x1, 1x1, 1x1, 2x2]. Note that, unlike the structure of the original VGG-19 model, no pooling layers are used in this work since pooling layers are known to cause information losses in the forward propagation of the network~\citep{2014arXiv1412.6806S} which are not desirable for the regression problem. Instead, the functionality of pooling layers are replaced by the strides. For the activation function between adjacent convolutional layers, \emph{exponential linear unit} function is used, while \emph{softmax} activation is used for the fully connected part of the network for classification. Dropout rate between layers is set as 0.01.

\subsection{Error Measurement in Training}
\label{apx:loss_func}

The successful training of VGG relies on whether the final output of the model returns the closest probability to the given type of GW, e.g., either lensed or unlensed signals in classification or the closest value to the target value in regression. In other words, we need to deploy an adequate error measurement method, $E$, (or, equivalently, loss function) to let the VGG can be properly updated through multiple iterations and, eventually, let it can return the desired output within acceptable tolerance on the error.

We use different forms of loss functions for the classification and regression separately. We compute the cross-entropy function:
\begin{equation}
E_\mathrm{c} = -p_i \log{\hat{p}_i}, \label{eqn:cross_entropy}
\end{equation}
for the classification at each iteration of training until the value of $E_\mathrm{c}$ satisfies given tolerance for the error. In Equation~(\ref{eqn:cross_entropy}), $\hat{p}_i$ and $p_i$ are the predicted value and the target value of an $i$-th training sample. Meanwhile, for the regression, we measure $E$ with mean squared error:
\begin{equation}
E_\mathrm{r} = \frac{1}{N} \sum^N_i ( \hat{p}_i - p_i )^2, \label{eqn:mse}
\end{equation}
where $N$ is the total number of training samples.

\section{Regression Results of $L_\mathrm{SIS}$}
\label{apx:reg_sis}

In this section, we present the regression results of $L_\mathrm{SIS}$: the residual plots (Figure~\ref{fig:reg_residual_sis}), the distributions of the true and predicted parameters (Figure~\ref{fig:reg_overlap_sis}), and the computed metrics, $\mathcal{M}$ and $\mathcal{D}_\mathrm{KL}$ (Table~\ref{tab:reg_metrics_sis}). One can see that all results are similar to the results of $L_\mathrm{PM}$ except $y$: $\mathcal{M}_y \simeq 0.956$ and $\mathcal{D}_{\mathrm{KL},y} \simeq 0.002$ (c.f. $\mathcal{M}_y \simeq0.876$ and $\mathcal{D}_{\mathrm{KL},y} \simeq0.051$ for $L_\mathrm{PM}$). Despite the result is notable, we just accept the exception in this paper and do not try to understand the reason because $y$ has the degeneracy depending on the randomly chosen physical parameters and, without the complete models on the physical parameters, more sophisticated systematical study on breaking the degeneracy in $y$ is rather meaningless at the moment.

\begin{table}[t]
\caption{The match, $\mathcal{M}$, and the Kullback-Leibler divergence, $\mathcal{D}_\mathrm{KL}$, between the true and predicted distributions of $M_S$, $M_L$, $z_S$, $z_L$, and $y$ of $L_\mathrm{SIS}$ depicted in Figure~\ref{fig:reg_overlap_sis}. Again, the bold faced values show the better result in the comparison between the source and lens systems. Like the result of $L_\mathrm{PM}$, $\mathcal{M}$ and $\mathcal{D}_\mathrm{KL}$ show consistent results that the prediction on the parameters of source system is better than those of lens system.\label{tab:reg_metrics_sis}}
\centering
\begin{tabular}{l | c c c c | c}
\hline
\hline
Metrics & $M_S^c$ & $M_L$ & $z_S$ & $z_L$ & $y$\\
\hline
$\mathcal{M}$ & \textbf{0.976} & 0.945 & \textbf{0.903} & 0.824 & 0.956\\
$\mathcal{D}_\mathrm{KL}$ & \textbf{0.0205} & 0.0207 & \textbf{0.051} & 0.111 & 0.002\\
\hline
\hline
\end{tabular}
\end{table}

\section{Regression Results of Lensing Parameters}
\label{apx:reg_magnification}

\begin{figure*}[t!]
    \centering
    \begin{turn}{-90}
    \begin{minipage}[c]{1.3\textwidth}
    \centering
    \subfigure[PM - $y$]
    {
        \includegraphics[width=.24\linewidth]{PRPlot_y_pm.pdf}
        \includegraphics[width=.24\linewidth]{PHist_y_pm.pdf}
    }
    \subfigure[SIS - $y$]
    {
        \includegraphics[width=.24\linewidth]{PRPlot_y_sis.pdf}
        \includegraphics[width=.24\linewidth]{PHist_y_sis.pdf}
    }\\
    \subfigure[PM - $\mu_{+}$]
    {
        \includegraphics[width=.24\linewidth]{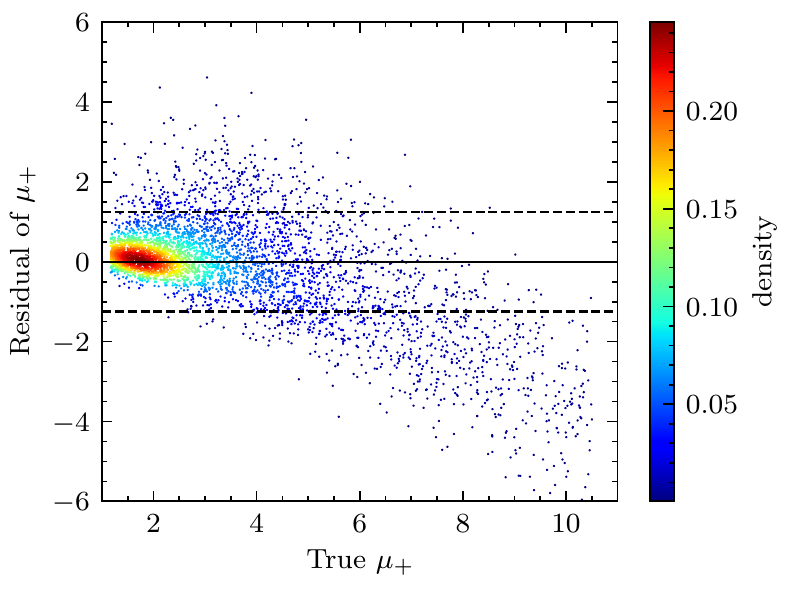}
        \includegraphics[width=.24\linewidth]{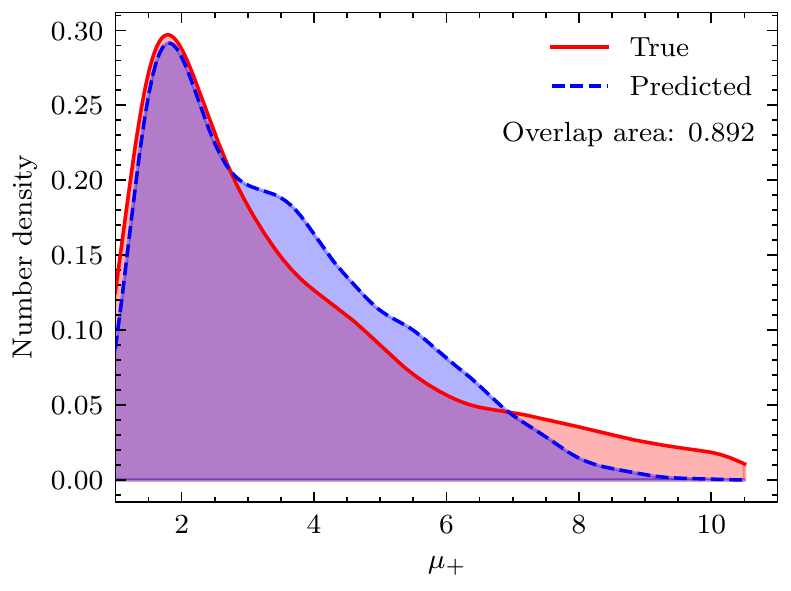}
    }
    \subfigure[SIS - $\mu_{+}$]
    {
        \includegraphics[width=.24\linewidth]{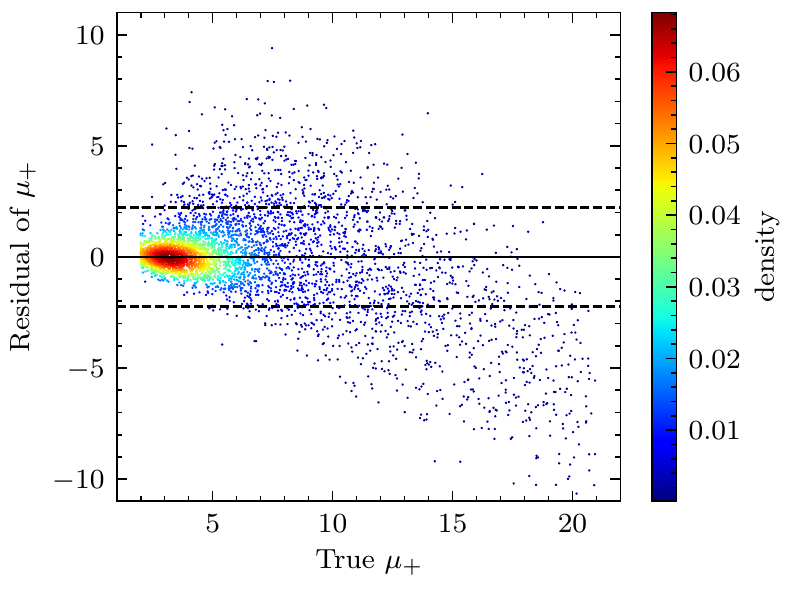}
        \includegraphics[width=.24\linewidth]{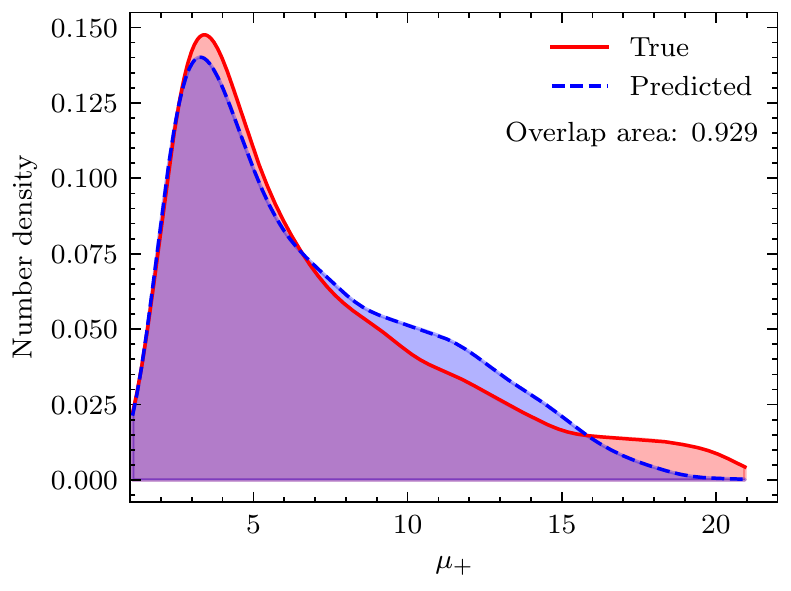}
    }\\
    \subfigure[PM - $\mu_{-}$]
    {
        \includegraphics[width=.24\linewidth]{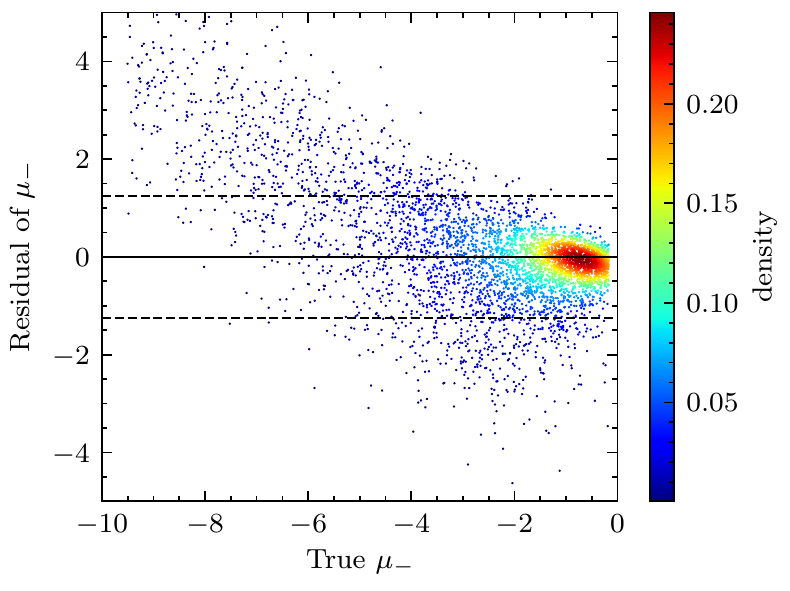}
        \includegraphics[width=.24\linewidth]{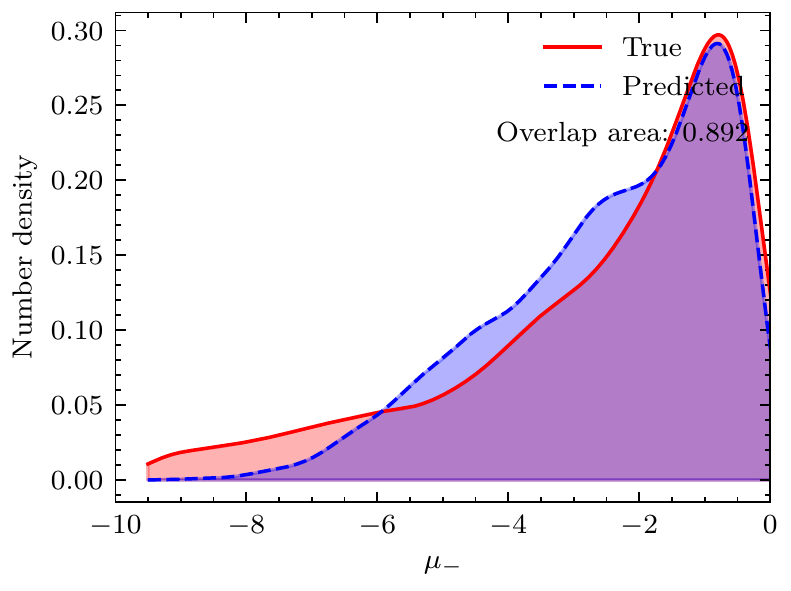}
    }
    \subfigure[SIS - $\mu_{-}$]
    {
        \includegraphics[width=.24\linewidth]{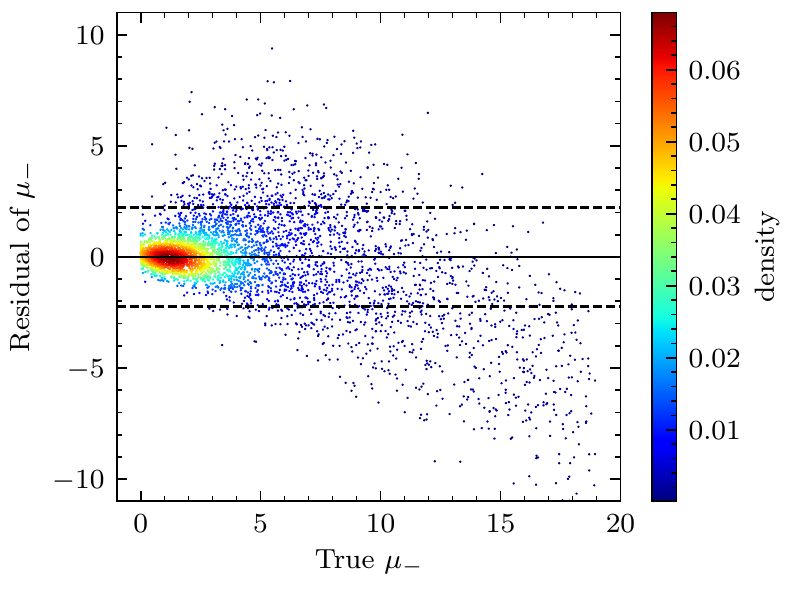}
        \includegraphics[width=.24\linewidth]{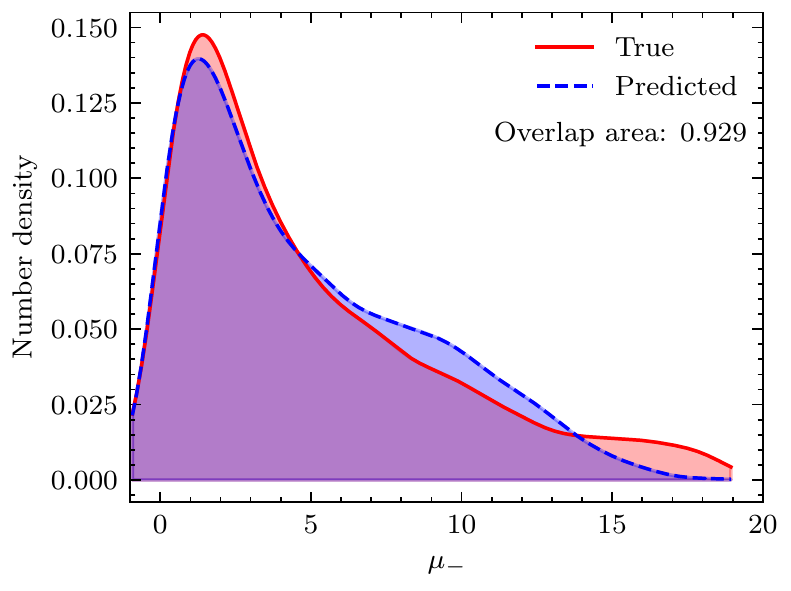}
    }
    \caption{Residual plots and distributions of $y$ and $\mu_\pm$ parameters of $L_\mathrm{PM}$ (left) and $L_\mathrm{SIS}$ (right). One can see that the residual plots and the distributions of predicted values of $\mu_\pm$ agree the relation between $y$ and $\mu_\pm$ given in Equations~\eqref{eq:mu_pm} and \eqref{eq:mu_sis} for PM and SIS, respectively. \label{fig:reg_lensing_params}}
    \end{minipage}
    \end{turn}
\end{figure*}

In this section, we present the result of additional experiments on the regression of another lensing parameter, the magnification factors, $\mu_\pm$. From the visual inspection of Figure~\ref{fig:reg_lensing_params}, we can see that the results agrees with the expected relations between $y$ and $\mu_\pm$ given in Equations~\eqref{eq:mu_pm} and \eqref{eq:mu_sis} for PM and SIS, respectively.

\section{Details of Microlensing Sample}
\label{apx:microlensing_params}

For the use of \texttt{lensingGW} in preparing a sample spectrogram of microlensed GW, we consider a simple case similar to the case discussed in Sec.~5.2.~of~\citet{Pagano:2020rwj}, that is, the microlensing of GWs enhanced by a macro lens like elliptic galaxy. We consider to set the parameters rather differently and arbitrarily such as $m_1=45 \msun$, $m_2=36 \msun$, $M_L=10^{12} \msun$, $z_S=0.2$, and $z_L=0.05$. With this setup, we get $\theta_E=2.65 \times 10^{-5}$ radian. The sky position (RA, Dec) of a source binary system is set as $(0.05, 0) \theta_E$ radian. The generation of an unlensed and non-precessing GW is done by using IMRPhenomPv2. For the distribution of microlenses, we assume that the point masses are distributed uniformly within the range of $[100, 200] \msun$ and their positions are uniformly distributed within $\pm 10^{-4} \times $(RA, Dec) around the (RA, Dec) of a chosen macro image. For this setup, \texttt{lensingGW} solver found 138 microlensed images. Finally, we convert the strain data into the spectrogram via the constant-Q transform method as done in this work. Because the setup is just prepared for testing and piloting purposes, we take a very high $\mathrm{SNR} \simeq 63$.

\bibliography{gw,astro,method,lens}
\bibliographystyle{aasjournal}

\end{document}